\documentclass{emulateapj}
\usepackage{color} 

\usepackage{epsfig}
\usepackage{subfigure}
\usepackage{graphicx}
\usepackage{amsmath}
\usepackage{natbib}
\newcommand{\pa}{\partial}
\newcommand{\mb}{\boldsymbol}
\newcommand{\wt}{\widetilde}

\shorttitle{Effect of Ambipolar Diffusion on MRI}
\shortauthors{Bai \& Stone}

\begin{document}


\title{Effect of Ambipolar Diffusion on the Non-linear Evolution of
Magnetorotational Instability in Weakly Ionized Disks}


\author{Xue-Ning Bai \& James M. Stone}
\affil{Department of Astrophysical Sciences, Princeton University,
Princeton, NJ, 08544} \email{xbai@astro.princeton.edu,
jstone@astro.princeton.edu}




\begin{abstract}
We study the role of ambipolar diffusion (AD) on the non-linear evolution of the MRI in
protoplanetary disks using the strong coupling limit, which applies when the electron
recombination time is much shorter than the orbital time. The effect of AD in this limit is
characterized by the dimensionless number $Am$, the frequency of which neutral
particles collide with ions normalized to the orbital frequency. We perform
three-dimensional unstratified shearing-box simulations of the MRI over a wide range
of $Am$ as well as different magnetic field strengths and geometries. The saturation
level of the MRI turbulence depends on the magnetic geometry and increases with the
net magnetic flux. There is an upper limit to the net flux for sustained turbulence,
corresponding to the requirement that the most unstable vertical wavelength be less
than the disk scale height. Correspondingly, at a given $Am$, there exists a maximum
value of the turbulent stress $\alpha_{\rm max}$. For $Am\lesssim1$, the largest stress
is associated with a field geometry that has both net vertical and toroidal flux. In this
case, we confirm the results of linear analyses that show the fastest growing mode has
a non-zero radial wave number with growth rate exceeding the pure vertical field case.
We find there is a very tight correlation between the turbulent stress $\alpha$ and the
plasma $\langle\beta\rangle\equiv P_{\rm gas}/P_{\rm mag}\approx1/2\alpha$ at the
saturated state of the MRI turbulence regardless of field geometry, and
$\alpha_{\rm max}$ rapidly decreases with decreasing $Am$. In particular, we find
$\alpha_{\rm max}\approx7\times10^{-3}$ for $Am=1$ and
$\alpha_{\rm max}\approx6\times10^{-4}$ for $Am=0.1$.

\end{abstract}



\section{Introduction}\label{sec:intro}

One of the most fundamental questions in accretion disk dynamics is how the disk
transports angular momentum and accretes to the central object. The magnetorotational 
instability (MRI, \citealp{BH91}) is widely considered to be the most likely mechanism for
the transport process.
The non-linear evolution of the MRI under ideal MHD conditions has been studied
extensively using both local \citep{HGB95,SHGB96,MillerStone00} and global
\citep{Armitage98,Hawley00,Hawley01,FromangNelson06} numerical simulations. It has
been found that MRI generates vigorous MHD turbulence and produce efficient outward
transport of angular momentum whose rate is compatible with observations.
However, accretion disks in some systems are only partially ionized, and non-ideal MHD
effects need to be taken into account. In particular, most regions of the protoplanetary
disks (PPDs) are too cold for sufficient thermal ionization, and effective ionization may be
achieved only in the disk surface layer due to external ionization sources such as X-ray
radiation from the central star and cosmic ray ionization \citep{Gammie96,Stone_etal00}.
Non-ideal MHD effects reflect the incomplete coupling between the disk material and the
magnetic field, and substantially affect the growth and saturation of the MRI.

There are three non-ideal MHD effects as manifested in the generalized Ohms's law,
namely the Ohmic resistivity, Hall effect and ambipolar diffusion (AD), with three
different regimes associated with the relative importance of these terms. In general, the
Ohmic term dominates at high density and very low ionization, the AD term dominates in
the opposite limit, and the Hall term is important in between. So far the majority of studies
have been focused on the Ohmic regime. In this case, MRI is damped when the Elsasser
number $\Lambda\equiv v_{\rm Az}^2/\eta\Omega$ falls below order unity
\citep{Jin96,Turner_etal07}, where $v_{\rm Az}$ is the Alfv\'en velocity in the
vertical direction, $\eta$ is the Ohmic resistivity, $\Omega$ is the angular frequency of
Keplerian rotation. Another often quoted criterion is the magnetic Reynolds number
Re$_M\equiv c_s^2/\eta\Omega\gtrsim10^4$ for MRI to be self-sustained (where $c_s$ is
the sound speed), which has the advantage of being independent of the magnetic field
\citep{Fleming_etal00}. Ohmic resistivity has been used extensively to model the layered
accretion in PPDs, where the surface layer of the disk is sufficiently ionized to couple to
the magnetic field and drive the MRI turbulence, while the midplane region is too poorly
ionized and ``dead"
\citep{FlemingStone03,Turner_etal07,TurnerSano08,IlgnerNelson08,OishiMacLow09}. 


The importance of Hall and AD terms in PPDs has been studied in a number of
theoretical works, but relatively little attention has been paid to numerical simulations
of the non-linear regime. Linear analysis of the MRI in the Hall regime have been
performed by \citet{Wardle99} and  \citet{BalbusTerquem01}. The growth rate is strongly
affected by the Hall term and depends on the sign of ${\mb B}\cdot{\mb\Omega}$.
Nevertheless, numerical simulations including both the Hall and Ohmic terms (where the
Ohmic term dominates) showed that the Hall term does not strongly affect the saturation
amplitude of MRI \citep{SanoStone02a,SanoStone02b}. It is yet to study the behavior of
MRI in the regime where Hall effect dominates over other terms, and to include the Hall
term in the more realistic vertically stratified simulations.

The relative motion between ions and neutrals leads to AD. AD is ideally studied
using the two-fluid approach, where the ions and neutrals are treated as separate fluids,
coupled by the ion-neutral drag via collisions. Moreover, ion and neutral densities are
affected by the ionization and recombination processes. Most analytical studies in the
linear regime adopt the Boussinesq approximation where ion and neutral densities are
kept constant \citep{BlaesBalbus94,KunzBalbus04,Desch04}. These studies show that
the growth of MRI is suppressed when the collision frequency of a neutral with the ions
falls below the orbital frequency. In the mean time, when both vertical and azimuthal field
is present, unstable modes always exist due to the effect of AD, and these unstable
modes require non-zero radial wavenumbers \citep{KunzBalbus04,Desch04}.
\citet{BlaesBalbus94} also studied the effect of ionization and recombination with
compressibility (for vertical propagating waves), and found that the presence of azimuthal
and radial field allows the coupling to acoustic and ionization modes, and the azimuthal
field tends to stabilize the flow when the recombination time is not too long compared
with dynamical time.

The effect AD on the MRI in the non-linear regime was first studied by \citet{MacLow_etal95}.
They implemented and tested AD in the ``strong coupling" limit (see below) in
the ZEUS code and performed simulations with net vertical flux for various ion-neutral
coupling strengths. Their results confirmed the stability analysis of \citet{BlaesBalbus94},
but their simulations are only two-dimensional and did not follow the evolution much
beyond the linear stage. In another study, \citet{Brandenburg_etal95} included the effect
of AD (also in the strong coupling limit) in their three-dimensional simulations of a local,
vertically stratified disk. They found that turbulence remains self-sustained in a case
where AD time is long compared with orbital time, although reduced in strength, and in
another case where the AD time was set comparable to $\Omega^{-1}$, turbulence
decayed.

A systematic study on the non-linear evolution of MRI with AD is done
by \citet{HawleyStone98} (hereafter HS) using three-dimensional (3D) numerical
simulations. They used the two-fluid approach without considering the
ionization-recombination processes, therefore ions and neutrals obey their own continuity
equations. Both net-vertical and net-toroidal magnetic configurations were considered.
They found that the system behaves like fully-ionized gas when the ion-neutral collision
frequency is greater than $100\Omega$, while ions and neutrals behave independently
when the collision frequency fall below $0.01\Omega$. The amplitude of magnetic field
at saturation is proportional to the ion density when it is much smaller than the neutral
density. The two-fluid approach adopted by HS is valid when the recombination time
scale is long compared with the dynamical time. However, AD in PPDs is in general in
the ``strong coupling" limit \citep{Shu91}. Two conditions must be satisfied in this limit:
1. The ion density $\rho_i$ is negligible compared with the neutral density $\rho_n$.
2. The electron recombination time $\tau_r$ must be much smaller than the orbital
frequency $\Omega$. In this limit, the ion density is purely determined by the
ionization-recombination equilibrium with the neutrals, and the two-fluid formulation is
simplified into a single-fluid formalism (for the neutrals). In PPDs, condition 1 is always
satisfied, and we will show in a companion paper \citep{Bai11} that condition
2 is almost always satisfied.

In this paper, we conduct three-dimensional (3D) local shearing-box simulations to
explore the effect of AD on the non-linear evolution of MRI in the strong coupling limit.
This is conceptually different from the simulations performed by HS in that the ion
density does not obey continuity equation, and is set by the neutral density due to
chemical equilibrium. Effectively, this allows the coupling of the MRI with acoustic and
ionization modes, which leads to more complicated interactions \citep{BlaesBalbus94}.
Moreover, our simulations correspond to the limit where the ion density is negligibly
small (i.e., $f\rightarrow0$ in HS), which is difficult for two-fluid simulations due to the
stiffness of the equations. In the strong coupling limit, there is only one controlling
parameter, namely, the ion-neutral collision frequency $\gamma\rho_i$. We perform
three sets of simulations with net vertical flux, net toroidal flux and both. In each group
of runs, we systematically vary $\gamma\rho_i$ as well as the strength of the net field.
Our main goal is to study the conditions under which MRI turbulence can be
self-sustained or is suppressed due to AD. In addition, we study the properties of the
MRI turbulence in the AD dominated regime.

This paper is structured as follows. In Section \ref{sec:formulation} we provide the
formulation of AD in the strong coupling limit, describe the numerical method and code
test problems. A series of numerical simulations on the non-linear evolution of MRI with
AD are presented and analyzed in Section \ref{sec:result}, from which we discuss the
condition under which MRI turbulence is sustained or suppressed as well as the
properties of the MRI turbulence in AD dominated regime. We conclude and briefly
discuss various implications In Section \ref{sec:conclusion}.


\section[]{Formulation and Numerical Method}\label{sec:formulation}

\subsection[]{Ambipolar Diffusion in Weakly Ionized Plasma}\label{ssec:nimhd}

In weakly ionized plasma, the inertia of the ionized species is negligible. The gas
dynamics can thus be described by single-fluid equations for the neutrals,
modified by non-ideal MHD effects. The effect of ambipolar diffusion (AD) derives from
the relative motion between the ions and the neutrals. Since the ion inertia is
negligible, the ion velocity is determined by the balance between the Lorentz force and
the ion-neutral collisional drag
\begin{equation}
\frac{{\mb J}\times{\mb B}}{n_i}=\gamma\rho m_i({\mb v}_i-{\mb v})\ ,
\end{equation}
where $m_i$, $n_i$ and ${\mb v}_i$ are the ion mass, number density and velocity
respectively, $\rho$, ${\mb v}$ are the density and velocity of the neutrals, ${\mb J}$,
${\mb B}$ are the current density and magnetic field vectors,
$\gamma=\langle\sigma v\rangle/(m_n+m_i)$, with $\langle\sigma v\rangle$ begin the
rate coefficient for momentum transfer between the ions and the neutrals and $m_n$
the neutral mass. The magnetic field is effectively carried by the ions, thus the induction
equation now reads
\begin{equation}
\begin{split}
\frac{\pa{\mb B}}{\pa t}&=\nabla\times({\mb v}\times{\mb B})
+\nabla\times\bigg[\frac{({\mb J}\times{\mb B})\times{\mb B}}
{c\gamma\rho\rho_i}\bigg]\\
&=\nabla\times({\mb v}\times{\mb B})
-\frac{4\pi}{c}\nabla\times\bigg(\frac{v_A^2}
{\gamma\rho_i}{\mb J}_\perp\bigg)\ ,
\end{split}\label{eq:induction1}
\end{equation}
where $\rho_i$ is the ion mass density, $v_A\equiv B/\sqrt{4\pi\rho}$ is the Alfv\'en
velocity, ${\mb J}_\perp$ is the component of the current density that is perpendicular
to the direction of the magnetic field. The second term on the right hand side is the
AD term, from which the ambipolar diffusivity can be defined as
$\eta_A=v_A^2/\gamma\rho_i$. The rest of the fluid equations remain the same
as the ideal MHD equations. In particular, the momentum equation is unchanged since
the neutrals effectively feel the magnetic tension and pressure by the ion-neutral
collisions.

The above derivation is strictly valid when electrons and ions are the only charge
carriers, and all ions have the same mass. Nevertheless, the effect of multiple ion
and charged grain species can be combined into an effective AD coefficient\footnote{A
more generalized derivation of all non-ideal MHD terms in the induction equation is
given and discussed in \citet{Bai11} (and see also \citet{Wardle99,Wardle07}), which
also includes Ohmic resistivity and the Hall term.}. AD becomes the dominant
non-ideal MHD effect when the gyro-frequency of both the ions and the electrons are
higher than their collision frequency with the neutrals (i.e., both ions and electrons are
coupled to the magnetic field). In practice, this corresponds to regions with low density
and strong magnetic field.

The effect of AD in rotating disks (with angular frequency $\Omega$) is characterized
by the parameter $Am$ \citep{CMC07}:
\begin{equation}
Am\equiv\frac{\gamma\rho_i}{\Omega}\ ,
\end{equation}
which is the number of effective collisions for a neutral molecule/atom with the ions
in a dynamical time $1/\Omega$. Physically, we see that $Am=v_A^2/\eta_A\Omega$,
which is equivalent to the Elsasser number for Ohmic resistivity
$\Lambda=v_A^2/\eta\Omega$. Therefore, $Am$ measures the ratio of the AD
time scale over the critical wavelength $v_A/\Omega$ and the dynamical time scale.

In this paper, we study AD in the strong coupling limit \citep{Shu91}: the electron (ion)
recombination time is much shorter than the dynamical time $1/\Omega$ so that the
ion density $\rho_i$ is determined by the local thermodynamical quantities (density
and temperature). The strong coupling limit is widely applicable in protoplanetary
disks, and we show in our companion paper \citep{Bai11} that the
recombination time is typically at least one order of magnitude smaller than the
dynamical time, even in the disk coronal regions. Using the strong coupling limit, we
assume the ion density depends on the neutral density $\rho$ in the form of
\begin{equation}
\rho_i=\rho_{i0}\bigg(\frac{\rho}{\rho_0}\bigg)^\nu\ ,\label{eq:rhoi}
\end{equation}
where $\rho_{i0}$ and $\rho_{0}$ are the reference density of the ions and the
neutrals. A simple-minded calculation of the ionization-recombination equilibrium
gives $\xi\rho=C\rho_i^2$, which yields $\nu=1/2$. In the equation $\xi$ is the
ionization rate, of the order $10^{-17}$ s$^{-1}$, $C$ is the effective rate coefficient
for recombination, of the order $10^{14}$ cm$^3$ s$^{-1}$ g$^{-1}$ in the absence
of grains \citep{BlaesBalbus94}. In reality, the recombination process is complicated
by a complex network of gas phase and grain phase chemical reactions, and we
can address the complications by exploring different values of $\nu$. Being a
two-body process in general, and one expects $0<\nu<1$.\footnote{For example,
Figure 1a and Figure 3 of \citet{BaiGoodman09} illustrate the dependence of
electron abundance $x_e\propto\rho_i$ on the gas density in various chemistry
models with and without dust grains, and $0<\nu<1$ holds in essentially all
circumstances.} In fact, our simulations show that the properties of the MRI turbulence
is insensitive to $\nu$ (see Section \ref{sssec:nu}).

\subsection[]{Numerical Method}\label{ssec:method}

We use Athena, a higher-order Godunov MHD code with constrained
transport technique to enforce the divergence-free constraint on the
magnetic field \citep{GardinerStone05,GardinerStone08,Stone_etal08}
for all calculations presented in this paper. Non-ideal MHD terms
including Ohmic resistivity \citep{Davis_etal10}, the Hall term (in progress)
and AD (this paper) have been developed for Athena.
We consider a local patch of a Keplerian disk using the standard
shearing-box formalism \citep{GoldreichLyndenBell65}, which adopts a
local reference frame at a fiducial radius corotating with the disk at orbital
frequency $\Omega$. In this frame, we write the MHD equations with AD
in a Cartesian coordinate system as
\begin{equation}\label{eq:gascont}
\frac{\pa\rho}{\pa t}+\nabla\cdot(\rho\mb{v})=0\ ,
\end{equation}
\begin{equation}
\frac{\pa\rho\mb{v}}{\pa t}+\nabla\cdot(\rho\mb{v}^T{\mb v}
+{\sf T})=\rho\bigg[2{\mb v}\times{\mb\Omega}
+3\Omega^2x\hat{\mb{x}}\bigg]\ ,
\label{eq:gasmotion}
\end{equation}
\begin{equation}
\frac{\pa{\mb B}}{\pa t}=\nabla\times\bigg[{\mb v}\times{\mb B}
+\frac{({\mb J}\times{\mb B})\times{\mb B}}
{c\gamma_i\rho_{i}\rho}\bigg]\ ,
\label{eq:induction2}
\end{equation}
where ${\sf T}$ is the total stress tensor
\begin{equation}
{\sf T}=(P+B^2/8\pi)\ {\sf I}-\frac{{\mb B}^T{\mb B}}{4\pi}\ ,
\end{equation}
${\sf I}$ is the identity tensor, $P$ is the gas pressure.
$\hat{\mb{x}},\hat{\mb{y}},\hat{\mb{z}}$ are
unit vectors pointing to the radial, azimuthal and vertical directions respectively, where
${\mb\Omega}$ is along the $\hat{\mb{z}}$ direction. Disk vertical gravity is ignored and
our simulations are vertically unstratified. We use an isothermal equation of state
$P=\rho c_s^2$, where $c_s$ is the isothermal sound speed. Periodic boundary
conditions are used in the azimuthal and vertical directions, while the radial boundary
conditions are shearing periodic as usual.

An orbital advection scheme \citep{FARGO,Johnson_etal08} has been implemented in
Athena \citep{StoneGardiner10}. It splits the dynamical equations into two systems, one
of which corresponds to linear advection operator with background flow velocity
$3\Omega x/2\hat{\mb y}$ and the other evolves only velocity fluctuations. The orbital
advection scheme not only accelerates the calculation in large box size by admitting
larger time steps, but also makes the calculation more accurate.

The last term on the right hand side of equation (\ref{eq:induction2}) represents
the AD term, where $\rho_i$ is the effective ion density, approximated by
equation (\ref{eq:rhoi}). Non-ideal MHD terms (e.g., the AD term) are implemented
in Athena in an operator-split way. Over one time step, one first solves the
induction equation using only the non-ideal MHD terms, and then solves ideal
MHD equations. The induction equation with AD term is a diffusion equation and is
evolved by a fully explicit forward-Euler method. The divergence free condition is
maintained using constrained transport: we calculate the AD electromotive force
defined at cell edges by interpolating the magnetic field and current density to such
locations. The method is stable, with time step constrained to be proportional to grid
size squared. The overall method is first order accurate in time.

Numerical method including the AD term has been frequently implemented in the
framework of two-fluid models (e.g., \citealp{Toth94,Stone97,SmithMacLow97,Falle03,
Li_etal06,OSullivanDownes06,OSullivanDownes07,TilleyBalsara08}), with the main
applications on star formation, turbulence and shocks in the interstellar medium.
However, single-fluid models for AD in the strong coupling limit relevant to weakly
ionized disks \citep{Brandenburg_etal95,MacLow_etal95} are relatively less studied
numerically. Recently, \citet{Choi_etal09} described an explicit scheme for
incorporating AD in the strong coupling limit in an MHD code that is second order
accurate, and use super time-stepping to accelerate the calculation when the AD
coefficient is large.
As we find MRI is suppressed at moderately large AD coefficients (see Section
\ref{sec:result}), super time-stepping is not needed for the purpose of this paper.

\subsection[]{Code Tests}\label{ssec:test}

In this subsection, we conduct two test problems to examine the performance of
our implementation of the AD term. For all these tests, we consider non-rotating
system and turn off the shearing box source terms on the right hand side of
equation (\ref{eq:gasmotion}).

\subsubsection[]{Isothermal C-type Shock Test}

The effect of ambipolar diffusion is best manifested in C-type shocks
\citep{Draine80}, which is a shock with continuous transitions consequent of the AD.
For the purpose of the code test, here we consider the isothermal C-type test by
\citet{MacLow_etal95}, which has become a standard test problem for AD.

We consider steady-state ($\pa/\pa t=0$) shock and work in the shock frame,
with upstream gas density $\rho_0$ moving at velocity $v_s$. The upstream gas
is threaded by a uniform magnetic field ${\mb B}_0$ that lies at an angle $\theta$
to the velocity. Let ${\mb v}_s$ be in the $\hat{x}$ direction, and ${\mb B}_0$ be
in the $\hat{x}-\hat{y}$ plane. For a continuous shock, the jump conditions reduce
to $\pa/\pa x=0$. Assuming the ion density $\rho_i$ is constant ($\nu=0$), the
equations that describe the C-type shock read
\begin{equation}
\rho v_x=\rho_0v_s\ ,
\end{equation}
\begin{equation}
\rho(v_x^2+c_s^2)+\frac{B_y^2}{8\pi}=\rho_0(v_s^2+c_s^2)
+\frac{B_{0y}^2}{8\pi}\ ,
\end{equation}
\begin{equation}
\rho v_xv_y-\frac{B_xB_y}{4\pi}=-\frac{B_xB_{0y}}{4\pi}\ ,
\end{equation}
\begin{equation}
v_xB_y-v_yB_x-\frac{B^2}{4\pi\rho\gamma\rho_i}
\frac{dB_y}{dx}=v_sB_{0y}\ .
\end{equation}
Note that $B_x=B_{0x}$ is constant.

\begin{figure*}
    \centering
    \includegraphics[width=150mm,height=70mm]{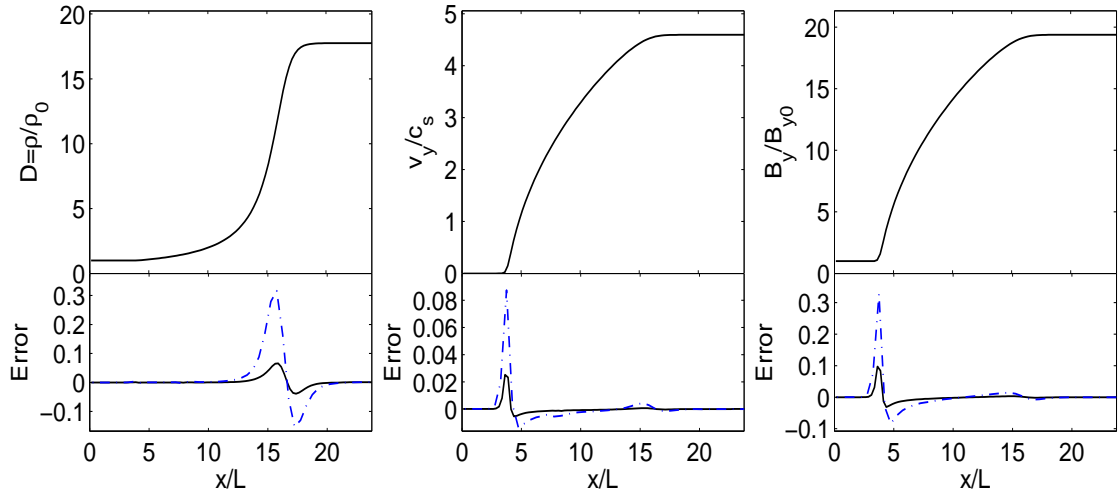}
  \caption{The profile of a C-type shock with $M=50$, $A=10$
  and $\theta=\pi/4$. The upper panels show the semi-analytical
  solution of gas density $\rho$, perpendicular velocity $v_y$ and
  perpendicular magnetic field $B_y$. $\rho$ and $B_y$ are
  normalized to their upstream values, and $v_y$ is normalized
  to the sound speed. The lower panels show the corresponding
  absolute errors (same units as the upper panels) from our numerical
  simulations, with resolutions of 4 cells per $L$ (black solid) and
  2 cells per $L$ (blue dash-dotted).}\label{fig:Cshock}
\end{figure*}

The shock is characterized by three dimensionless parameters: the
sonic Mach number $M=v_s/c_s$, the Alfv\'en Mach number $A=v_s/v_A$ (where
$v_A^2=B_0^2/4\pi\rho_0$), and the angle $\theta$ of the magnetic field with the
upstream flow. The characteristic length scale of the problem is given by
$L=v_A/\gamma\rho_i$. We further define $D\equiv\rho/\rho_0$, and $b\equiv B_y/B_0$.
After some algebra, we arrive at a dimensionless first order differential equation for $D$
\citep{MacLow_etal95}
\begin{equation}
\begin{split}
\bigg(\frac{1}{D^2}-&\frac{1}{M^2}\bigg)\frac{dD}{d(x/L)}
=(b^2+\cos^2\theta)^{-1}\times\\
&\times\frac{b}{A}\bigg[b-D\bigg(\frac{b-\sin\theta}{A^2}\cos^2\theta
+\sin\theta\bigg)\bigg]
\ .\label{eq:Cshock}
\end{split}
\end{equation}
One can numerically integrate this ordinary differential equation to obtain the C-type
shock profile. In Figure \ref{fig:Cshock} we show a semi-analytical solution for
$M=50$, $A=10$ and $\theta=\pi/4$ obtained by using a 4th order Runge-Kutta method. 

To use this solution as a code test, the shock is set to be aligned with the grid in the
$\hat{\mb x}$ direction. We use outflow boundary conditions in this direction. In
multi-dimensional tests, periodic boundary conditions are used in other directions.
The shock solution should be stationary (i.e., a standing shock), thus we evolve
the solution for sufficiently long time ($\sim5L/c_s$) and compare to the initial conditions.
In Figure \ref{fig:Cshock}, we further show the absolute error of the shock
profile compared with the semi-analytic solution. Since the shock is
grid-aligned, 1D, 2D and 3D tests essentially produce the same result. In our tests,
the grid resolution is chosen to be 2 and 4 cells per $L$.\footnote{In comparison,
\citet{MacLow_etal95} achieved comparable accuracy as ours using 5 and 10 cells
per $L$, \citet{Choi_etal09} achieved similar or better accuracy at 6.4 cells per $L$.}
We see that our code very accurately resolves the structure of the C-type shock using
only a few cells per $L$. The main source of the error lie in the region where density and
velocity profile vary quickly. In our comparison, the position of the shock is fixed at
the initial place, while in reality, the shock position can shift slightly during numerical
relaxation.

\subsubsection[]{Damping of MHD Waves}

Linear MHD waves are damped due to AD. Because the exact eigenvectors in the
AD regime can be very complicated, we initialize the problem with ideal MHD wave
eigenvectors and measure the damping rate. This means the initial conditions are
a linear superposition of more than one eigen-mode,
but the averaged damping rate should approach the analytical value for a single mode
as long as the AD coefficient is sufficiently small.

The analytical damping rate for various MHD waves due to AD can be found in
and/or derived from \citet{Balsara96}, as we summarize below. The damping rate of
the Alfv\'en wave is given by the solution of
\begin{equation}
\omega^2=k^2v_A^2\cos^2\theta
\bigg(1-{\rm i}\frac{\omega}{\omega_a}\bigg)\ .
\end{equation}
where $\omega_a\equiv\gamma\rho_i$, $v_A$ is the Alfv\'en velocity, $\theta$ is the
angle between the magnetic field and the wave vector. The damping rate of fast and
slow waves can be obtained by solving the quadratic equation
\begin{equation}
(\omega^2-v_f^2)(\omega^2-v_s^2)+{\rm i}
(\omega^2-k^2c_s^2)k^2v_A^2\frac{\omega}{\omega_a}=0\ ,
\end{equation}
where $v_f$ and $v_s$ are the fast and slow magnetosonic speeds.

\begin{figure*}
    \centering
    \includegraphics[width=150mm,height=65mm]{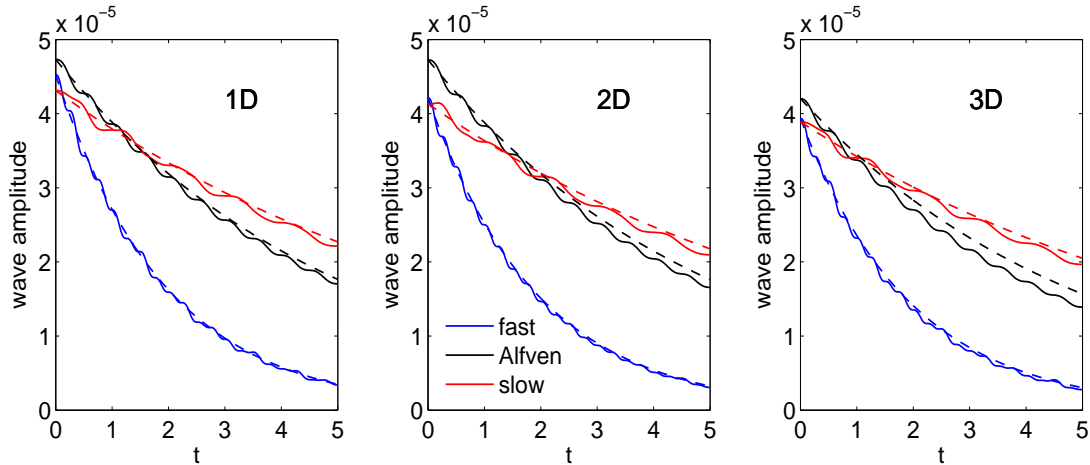}
  \caption{The damping of linear MHD waves by ambipolar diffusion.
  Three panels from left to right show the test results for 1D, 2D and
  3D, where in the latter two cases the waves are not grid-aligned.
  In each panel, black, blue and red curves show the damping of
  Alfv\'ven, fast and slow waves respectively. Solid lines are the
  measured damping curve, while dashed lines are the expected
  damping curve.}\label{fig:ADDamp}
\end{figure*}

When $\omega\ll\omega_a$, the damping rate is small and can be found by expanding
the ideal MHD dispersion relation to powers of $\omega/\omega_a$, and to the first
order, we find for the damping rate of the Alfv\'ven wave
\begin{equation}
\Gamma_A=\frac{1}{2}\frac{k^2v_A^2\cos^2\theta}{\omega_a}\ .
\end{equation}
The damping rate for fast and slow magnetosonic waves are
\begin{equation}
\Gamma_f=\frac{1}{2}\bigg(\frac{v_f^2-c_s^2}{v_f^2-v_s^2}\bigg)
\frac{k^2v_A^2}{\omega_a}\ ,\qquad
\Gamma_s=\frac{1}{2}\bigg(\frac{c_s^2-v_s^2}{v_f^2-v_s^2}\bigg)
\frac{k^2v_A^2}{\omega_a}\ .
\end{equation}

We perform the linear wave damping test in 1D, 2D and 3D. In 1D, the wave is
grid-aligned, whose wave length equals $1$ in code unit. In 2D and 3D test
problems, the wave vectors are not grid-aligned, with box sizes chosen such that
the wave length is also $1$ [in 2D, the box size is ($\sqrt{5}, \sqrt{5}/2$) and in 3D,
the box size is (3, 1.5, 1.5)]. We use isothermal equation of state with $c_s=1$.
The background gas density is $\rho_0=1$. As before, we choose $\nu=0$.
In 1D, the wave vector is along the $x$ direction, and the adopted magnetic field is
$B_{0x}=1.0$, $B_{0y}=\sqrt{2}$ and $B_{0z}=0.5$. In 2D and 3D the background
magnetic field vector is rotated with the wave vector accordingly while keeping
$\theta$ the same as in 1D (and a vector potential is used to initialize the wave in
order to preserve the divergence free condition). From the above we get
$v_A=\sqrt{13/4}$, $v_f=2$ and $v_s=1/2$, therefore we find
$\Gamma_A=2\pi^2Q_A$, $\Gamma_f=5.2\pi^2Q_A$ and
$\Gamma_s=1.3\pi^2Q_A$.

In practice, we adopt $\omega_a=100$  and run our simulation to $t=5$. By
default, the grid size is $32$ in 1D, $64\times32$ in 2D, and $64\times32\times32$
in 3D. Accounting for the box size in each dimension, the effective resolution,
characterized by number of cells per wavelength, is $32$, $28.6$ and $21.3$ in
1D, 2D and 3D respectively. The results are shown in Figure \ref{fig:ADDamp}.
From left to right, we show the damping curve from 1D, 2D and 3D simulations in
the solid lines, where black, blue and red lines label Alfv\'en, fast and slow MHD
waves. Dashed lines show the theoretical damping curve. We see that the
numerical damping rate matches very well with the theoretical damping rate. In the
3D runs, the damping rate for the Alfv\'en wave is slightly faster than expected, but
this may be because the effective resolution is less.

We have also run the simulations with double and half the resolution. With double
resolution, the numerical damping curves in 1D, 2D and 3D cases almost match
exactly the analytical damping curves (besides some small oscillations due to the
initial conditions). At half resolution, however, the numerical damping rate deviates
substantially (about $15\%$ to $30\%$ at $t=5$). These results indicate that at
least $20$ cells per wavelength is need to accurately capture the effect of AD.

\section[]{Simulations and Results}\label{sec:result}

In this section we describe three groups of simulations and study the effect of
AD on the non-linear evolution of the MRI. All our simulations are vertically
unstratified by ignoring the disk vertical gravity with fixed box height to be one
disk scale height $H=c_s/\Omega$. We initialize our simulations with Keplerian
velocity and seed density perturbations of $2.5\%$ of the background density
$\rho_0$. We consider three different magnetic field geometries (net vertical flux,
net toroidal flux, both vertical and toroidal flux) as described in the following
three subsections. Since all our simulations contain net magnetic flux, they are
not subject to the issue of convergence found by \citep{FromangPap07b} in zero
net-flux simulations, and numerical convergence is confirmed in our test
simulations (and see Section \ref{sssec:convergence} for the case of net toroidal
flux). For relatively small AD coefficient (large $Am$), MRI grows
quickly from the seed perturbations and saturates into turbulence; when the
effect of AD is strong, however, MRI does not grow from our seed perturbations.
In such cases, we initialize the simulations from a turbulent state which is
obtained from simulations with relatively large $Am$ (see individual subsections
for details).

The most important diagnostics are the volume averaged (normalized) Reynolds
stress, defined as
\begin{equation}
\alpha_{\rm Re}=\frac{\overline{\rho v_xv'_y}}{\rho_0c_s^2}\ ,
\end{equation}
where the over bar indicates volume averaging, $v'_y$ is the azimuthal velocity
with the Keplerian velocity subtracted, and the volume averaged (normalized)
Maxwell stress, defined as
\begin{equation}
\alpha_{\rm Max}=\frac{\overline{-B_xB_y}}{4\pi\rho_0c_s^2}\ .
\end{equation}
The total stress, namely, the $\alpha$ parameter \citep{ShakuraSunyaev73} is
$\alpha=\alpha_{\rm Re}+\alpha_{\rm Max}$. We also monitor the kinetic and
magnetic energy density, which characterize the strength of the MRI turbulence. 

The main purpose of this study is to identify the criterion under which sustained
turbulence generated by the MRI can be maintained. However, the term
``sustained turbulence" is a somewhat ambiguous concept. In the context of 3D
shearing box simulations, small-amplitude oscillations left from decayed
hydrodynamical turbulence is present \citep{ShenStone06}. Such oscillations
produce Reynolds stress on the order of $10^{-4}$ with kinetic energy density on
the order of $10^{-3}$, both in normalized unit $\rho_0c_s^2$. Such oscillations are
likely to associate with linear modes in the shearing sheet with the origin of
acoustic and/or nearly incompressible inertia waves \citep{Balbus03}.  Being a
conservative Godunov MHD code, the Athena code preserves the amplitude of
these waves without much damping. Therefore, throughout this paper, the level
of turbulence we are interested in are those whose time and volume averaged
kinetic energy density $E_k=\langle\rho v^2\rangle$ is on the order of
$10^{-3}\rho_0c_s^2$ or higher, and/or whose total stress $\alpha$ is no less than
$10^{-4}$. Meanwhile, analysis of all our simulations show that the threshold where
the MRI turbulence can be marginally self-sustained is roughly at the same level.
(see Section \ref{sssec:criterion} for further discussion).

Our simulations are run for at least $24$ orbits ($150\Omega^{-1}$). A period of $24$
orbits is sufficiently long for the MRI to saturate from initial growth, which typically occurs
in $5-10$ orbits, or for the restart runs to reach a steady state, which typically occurs in
$10-15$ orbits. Our time averaged quantities are mostly taken from after about $16$ orbits
(since $100\Omega^{-1}$) unless otherwise noted. Although a time average over $8$
orbits ($50\Omega^{-1}$) is relatively short, it is sufficient for our purpose to judge whether
MRI turbulence can be self-sustained\footnote{\citet{Winters_etal03} found that more than
a few hundred orbits as are required to accurately measure the properties of the MRI
turbulence in ideal MHD. This conclusion is based simulations with radial box size being
$H$, while the radial box size in about half of our simulations is $4H$, which reduces the
time fluctuations. Also, our Figures \ref{fig:alphBzhist} and \ref{fig:alphByhist} show that the
fluctuations in the Maxwell stress are less severe in the presence of AD than in the ideal
MHD case.}. Many of our simulations are run for $48$ orbits or longer where better statistics
on the turbulence properties can be obtained.

\subsection[]{Net Vertical Flux Simulations}\label{ssec:netBz}

In the first group of simulations, we choose the initial field configuration to be uniform
along the vertical axis $\hat{z}$, characterized by the plasma $\beta_0=2P_0/B_0^2$,
where $P_0=\rho_0c_s^2$ is the background pressure and $B_0$ is the initial field
strength. The vertical flux is conserved numerically by remapping the toroidal
component of the magnetic field in the ghost zones of the radial boundaries (see
Section 4 of \citet{StoneGardiner10} for details). For all simulations, we fix the box size
to be $4H\times4H\times H$ in the radial, azimuthal and vertical dimensions, with
fixed grid resolution at $64$ cells per $H$. We have chosen a relatively large radial
box size ($4H$), as suggested by \citet{PessahGoodman09}, which is needed to
properly capture the parasitic modes to break the channel mode into turbulence. It also
help substantially reduce the intermittence of the MRI turbulence \citep{Bodo_etal08}.
We note that for local unstratified net vertical flux MRI simulations without explicit
dissipation, turbulence properties converge at about $32$ cells per $H$ \citep{HGB95}.
The grid resolution in our simulations is two times higher, thus we expect  numerical
convergence.

All our net vertical flux simulations are listed in Table \ref{tab:netBz}. We first perform
a fiducial set of simulations with fixed $\beta_0=400$. We choose a series of $Am$ values,
ranging from $1000$ down to $0.1$, and study the critical value of $Am$ below which
MRI turbulence is no longer self-sustained (Section \ref{sssec:fidBz}). In the next, we
vary the net vertical flux by setting $\beta_0=100, 1600$ and $10^4$ and run a number of
simulations around $Am=1$ to study how the critical value of $Am$ is affected by the
vertical flux (Section \ref{sssec:strBz}). Moreover, in Section \ref{sssec:nu} we briefly
investigate the effect of $\nu$ by varying $\nu$ from the fiducial value $0.5$ to $0$ (run
Z5a) and $1$ (run Z5b) (see equation (\ref{eq:rhoi})). Finally, we discuss the properties
of the MRI turbulence in the presence of AD (Section \ref{sssec:turbprop}).

Our choices of the net vertical flux derive from the linear dispersion relation of the MRI
as well as physical considerations. In the case of ideal MHD, the wavelength for the
fastest growing linear MRI mode is given by $\lambda/H=9.18 \beta_0^{-1/2}$ \citep{HGB95}.
For $\beta_0=100, 400$ and $1600$, our vertical box size of $H$ fits $1, 2$ and $4$ most
unstable wavelengths respectively in ideal MHD. The ideal MHD dispersion relation is
considerably modified when $Am\lesssim10$. Unstable modes exist for wavelength longer
than the critical wavelength \citep{Wardle99}
\begin{equation}
\frac{\lambda_c}{H}=5.13\bigg(1+\frac{1}{Am^2}\bigg)^{1/2}\beta_0^{-1/2}\ .\label{eq:disp_u}
\end{equation}
The wavelength for the most unstable mode $\lambda_m$ is about twice larger. An
approximate fitting formula that is accurate within $2\%$ for all values of $Am$ is
\begin{equation}
\frac{\lambda_m}{H}\approx10.26\bigg(1+\frac{1}{Am^2}+\frac{1}{Am^{1.16\epsilon}}
-0.2\epsilon\bigg)^{1/2}\beta_0^{-1/2}\ ,\label{eq:disp_m}
\end{equation}
where $\epsilon\equiv Am/(1+Am)$. Note that for pure vertical magnetic field and vertical
wavenumber, the linear dispersion relation for Ohmic resistivity is exactly the same as that
for AD \citep{Wardle99}, with $Am$ replaced by the Elsasser number
$\Lambda=v_{Az}^2/\eta\Omega$. For $\beta_0=400$, the most unstable wavelengths at
$Am=1, 1/3$ and $0.1$ are $\lambda=0.87H$, $1.72H$ and $5.18H$ respectively. Clearly,
the most unstable mode does not fit into our simulation box when $Am=0.33$, and no
unstable modes fit into the box for $Am=0.1$. Since $\lambda\propto\beta_0^{-1/2}$, these
modes do fit into our simulation box as we increase $\beta_0$ to $1600$ and $10^4$
respectively. In the mean time, since AD tends to be important in the more strongly
magnetized upper layers of the protoplanetary disks \citep{Wardle07,Bai11}, it is
also interesting to study whether the MRI turbulence can be sustained when $\beta_0$ is
relatively small, even if the most (or all) unstable modes do not fit into our simulation box.
We have not run simulations with a taller box since we do not include vertical stratification.

\begin{table}
\caption{Net vertical flux simulations.}\label{tab:netBz}
\begin{center}
\begin{tabular}{cccccccc}\hline\hline
 Run & $Am$ & $\beta_0$ & $\nu$ & Orbits & Restart$^1$ & Turbulence$^2$ \\\hline
Z1 & 1000 & 400 & 0.5 & 48 & No & Yes \\
Z2 & 100   & 400 & 0.5 & 24 & No & Yes \\
Z3 & 10     & 400 & 0.5 & 24 & No & Yes \\
Z4 & 3.33  & 400 & 0.5 & 24 & No & Yes \\
Z5 & 1       & 400 & 0.5 & 43 & No & Yes \\
Z6 & 0.33  & 400 & 0.5 & 24 & Z5 & No \\
Z7 & 0.1     & 400 & 0.5 & 24 & Z5 & No \\
Z3s & 10   & 100 & 0.5 & 24 & No & Yes \\
Z5s & 1     & 100 & 0.5 & 24 & No & Yes \\
Z6s & 0.33  & 100 & 0.5 & 24 & Z5s & No \\
Z7s & 0.1   & 100 & 0.5 & 24 & Z5s & No \\
Z3w & 10  & 1600 & 0.5 & 24 & No & Yes \\
Z5w & 1     & 1600 & 0.5 & 24 & No & Yes \\
Z6w & 0.33 & 1600 & 0.5 & 48 & No & Yes \\
Z7w & 0.1   & 1600 & 0.5 & 24 & Z5w & No \\
Z6e & 0.33  & $10^4$ & 0.5 & 48 & No & Yes \\
Z7e & 0.1   & $10^4$ & 0.5 & 48 & Z6e & Yes \\
Z5a & 1     & 400 & 0.0 & 24 & No & Yes \\
Z5b & 1     & 400 & 1.0 & 24 & No & Yes \\
\hline\hline
\end{tabular}
\end{center}
Box size is fixed at $4H\times4H\times H$, grid resolution
is $64$ cells per $H$.

$^1$ whether simulation is initiated by restarting from a turbulent run.

$^2$ whether turbulence can be self-sustained.
\end{table}

\subsubsection[]{A Fiducial Set of Runs}\label{sssec:fidBz}

\begin{figure}
    \centering
    \includegraphics[width=92.5mm]{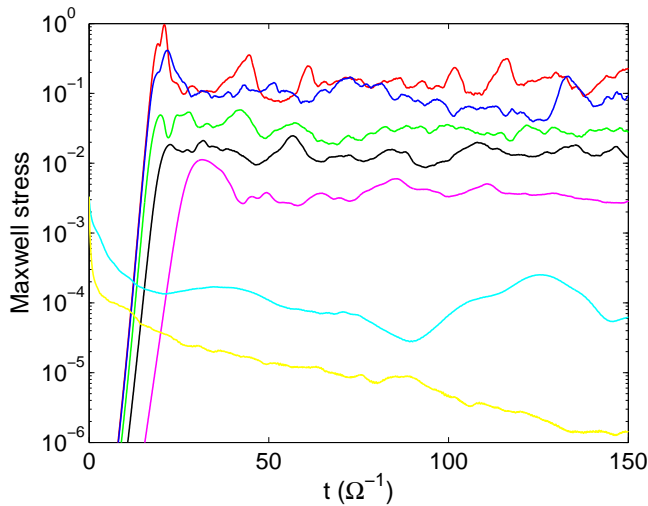}
  \caption{The evolution of Maxwell stress in our fiducial set of net vertical flux
  runs (from top to bottom: Z1, Z2, ..., Z7) normalized to $c_sH$. For models Z6 and Z7,
  the simulations are initiated from the end of run Z5.}\label{fig:alphBzhist}
\end{figure}

As the fiducial set of runs, we fix $\beta_0=400$, and run $7$ simulations with different
$Am$ values (see Table \ref{tab:netBz}), labeled from $Z1$ with $Am=1000$, which
essentially corresponds to the ideal MHD case, to $Z7$ with $Am=0.1$, where the
evolution of magnetic field is dominated by AD. Our scan of $Am$ is more narrowly
sampled near $Am=1$ where the transition is expected to occur. In Figure
\ref{fig:alphBzhist}, we show the time evolution of the Maxwell stress from the fiducial
set of runs. We find that for $Am\geq1$, the growth of the MRI from linear perturbations
leads to vigorous MRI turbulence, while for the two runs with $Am<1$, MRI either does
not grow from the initial vertical field (Z7), or grows too slowly (Z6), since the most
unstable modes do not fit into our simulation box. Therefore, for these two models, we
start the simulations from the end of run Z5 ($Am=1$), which is turbulent, and reset
$Am$ to be $0.33$ and $0.1$ respectively. Nevertheless, turbulence continues to
decay throughout the span of our simulation in run Z7. Run Z6 is a marginal case
where turbulence is neither fully sustained nor decayed continuously (see discussions
below).

We first look at run Z1 to Z5 with $Am\geq1$. The initial growth of the MRI is due to the
axisymmetric channel mode \citep{GoodmanXu94,PessahChan08}. The mode
becomes non-linear (producing an overshoot in the Maxwell stress up to $1$ in the
ideal MHD case) until broken down by secondary parasitic modes to produce turbulence.
In the turbulent state, it is evident that the Maxwell stress monotonically decreases as $Am$
decreases, analogous to the Ohmic case \citep{Sano_etal98,Fleming_etal00}. In Table
\ref{tab:turbBz} we list the general properties of the turbulence from all our vertical net flux
simulations. The quantities are averaged over space and time after saturation (after
$100\Omega^{-1}$). The total stress $\alpha\approx0.2$ in run Z1, which agrees with the
ideal MHD case \citep{HGB95}. It drops slowly with decreasing $Am$ when $Am\gg1$,
but very rapidly when $Am$ is around 1. Moreover, as $Am$ decreases, the ratio of kinetic
to the fluctuating part (with background field $B_{z0}$ subtracted) of the magnetic
energy increases (see also Figure \ref{fig:pwr_Bz}). Similarly, the ratio of Reynolds
stress to Maxwell stress increases.


As we have discussed before, the most unstable mode does not fit into our simulation
box for runs Z6 and Z7, and our simulations initiated from a turbulent state also show no
sign of sustained MRI turbulence. This is not surprising for run Z7, where no unstable MRI
mode even exists in the simulation box.
Our run Z6 is a marginal case, where a wavelength of $H$ is only slightly larger than the
critical wavelength for instability ($\lambda_c=0.89H$) but far from the most unstable
wavelength ($\lambda_m=1.72H$). This explains the long-term variations in the Figure
\ref{fig:alphBzhist} since the growth rate is only slightly larger than zero. Our analysis in
Section \ref{sssec:criterion} indicates that although some non-zero stress close to
$\alpha=10^{-4}$ is maintained in the simulation, it is unlikely to be due to the MRI
turbulence. In real disks, one may expect sustained turbulence to be supported at
$Am\approx0.3$ if it were at the disk midplane, where the density variation over one $H$
above and below the midplane is not significant. In the upper layers, $\beta_0$ may fall off
substantially over one $H$, which strongly stabilizes the flow. In sum, we see from our
fiducial set of net vertical flux simulations that presence of turbulence mainly depends on
whether the most unstable mode of the MRI fits into the simulation box. This aspect will
be further explored in the next subsection \ref{sssec:strBz}.

The results reported above are qualitatively different from those observed in
HS using two-fluid simulations. One may compare our results with Table 3 of HS,
where the $Am$ value for their four runs Z24, Z17, Z25, Z28 are $0.11$, $1.1$,
$11.1$ and $111$ respectively. The total stress $\alpha$ in these simulations does
not scale monotonically with $Am$, and in particular $\alpha$ is on the order of
$10^{-2}$ when $Am$ is as small as $0.11$, while in our simulations MRI is
suppressed. This reflects the difference in the physical assumptions about the two
approaches. In the two-fluid limit, ions are coupled to the neutrals only via collisions.
When the ion-neutral collisions are infrequent ($Am<1$), the ions and neutrals
behave as  independent fluid: vigorous MRI is generated in the ion fluid while the
neutrals remain quiescent, and the overall $\alpha$ is proportional to the ion fraction
$f=\rho_i/\rho$. In the strong-coupling limit, the ions and the neutrals are coupled
not only by collisions, but also via the ionization-recombination reactions. Since
the strong-coupling limit requires the recombination time to be much smaller than
the orbital time, this means that ions are continuously created and destroyed on a
time scale that is much shorter than the time scale for MRI to grow. It is this
additional chemical coupling that quenches the MRI in the ion fluid, and suppresses
angular momentum transport for $Am\lesssim0.1$.

\subsubsection[]{The Effect of Vertical Field Strength}\label{sssec:strBz}

We select a number of models from the fiducial series and rerun the simulations
with three additional initial $\beta_0$ values: $\beta_0=100, 1600$ and $10^4$ (e.g., with
magnetic field strength two times and half of that in the fiducial models, as well as one
case with an very weak field). These simulations are labeled with an additional letter
``s" (for strong), ``w" (for weak) and ``e" (for extremely weak) in Table \ref{tab:netBz}.
For strong field simulations with $\beta_0=100$, the wavelength for the fastest growing
mode exceeds the vertical box size when $Am<10$. When $Am<1$, there are
essentially no unstable mode in the simulation box. Runs Z3s and Z5s are initiated
from seed perturbations, while runs Z6s and Z7s are initiated from the turbulent state at
the end of run Z5s. For weak field runs with $\beta_0=1600$, on the other hand, the
most unstable mode can be fitted into the simulation box for all runs with
$Am\gtrsim0.3$.  No unstable mode is fitted into the simulation box when $Am=0.1$,
and as before, Run Z7w is initiated from turbulent state from Z5w to test whether
turbulence can be sustained. Our $\beta_0=10^4$ simulations allows the most unstable
wavelength to be fitted in the simulation box at small $Am$. We conduct two runs in this
case with $Am=0.33$ (Z6e) and $Am=0.1$ (Z7e). Run Z7e is initialized from the
turbulent state in Z6e to avoid the extremely long time in the linear growth stage. Time
averaging in runs Z6w, Z6e and Z7e are taken since $t=200\Omega^{-1}$ (time averaging
in other runs are taken since $t=100\Omega^{-1}$  by default).

We find from our simulations that sustained MRI turbulence is present in all models
except Z6s, Z7s and Z7w. In particular, the MRI turbulence can be self-sustained
even the $Am$ value is as small as $0.1$, provided that the net vertical field is
sufficiently weak. These results confirm our speculation in the fiducial set of
simulations that the MRI turbulence is self-sustained as long as unstable MRI modes
fit into the simulation box.

The diagnostic quantities from time and volume averaged quantities in the turbulent
state from the weak and strong field series of runs are also listed in Table
\ref{tab:turbBz}. We see that for $Am=10$, the averaged kinetic energy, Reynolds
and Maxwell stress monotonically decreases with increasing $\beta_0$. Although not
all our simulations are not run long enough for these quantities to be measured
accurately, the trend is significant enough and indicates that the MRI
saturate at a higher level with higher net vertical flux (small $\beta_0$), in agreement
with the ideal MHD case \citep{HGB95}. For $Am=1$, the monotonicity trend is still
present by comparing our fiducial run Z5 and the weak field run Z5w. The saturation
level of the MRI turbulence in the strong field run Z5s is weaker than that for run Z5.
This is most likely because the most unstable mode does not fit into our simulation
box (but some less unstable modes fit) in run Z5s. The monotonicity trend further
preserves at $Am=0.33$, where the kinetic energy density and total stress from run
Z6w is larger than those from run Z6e by about a factor of 2.

\begin{figure}
    \centering
    \includegraphics[width=92.5mm,height=75mm]{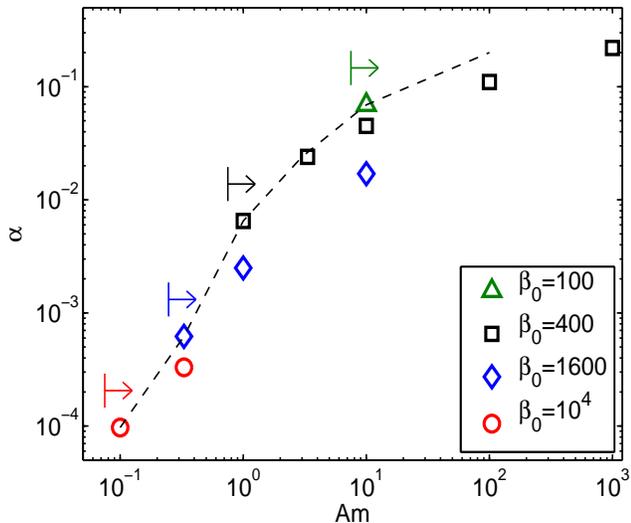}
  \caption{The time and volume averaged total stress $\alpha$ from all our net
  vertical flux simulations that sustain MRI turbulence. Simulations with different
  net vertical flux, characterized by the plasma $\beta_0$, are labeled by different
  symbols and colors. The arrows above the symbols indicate (for each $\beta_0$
  as represented by the symbol) the range of $Am$ where the most unstable
  wavelength is smaller than $H$. The dashed line connecting the symbols
  represents the maximum value of stress attainable from net vertical flux
  simulations.}\label{fig:alphBzall}
\end{figure}

We summarize the main results from the net vertical flux simulations in Figure
\ref{fig:alphBzall}. Shown are the total stress $\alpha$ from all the simulations that
the most unstable mode is properly resolved so that the MRI turbulence is
self-sustained and a reliable value of $\alpha$ can be obtained. As discussed before,
at a fixed $\beta_0$, there exists a critical value of $Am$ below which the most
unstable wavelength would exceed $H$ and the mode tend to be suppressed due to
the vertical stratification. This is effect is illustrated by the colored arrows in the Figure.
Equivalently, for turbulence to be sustained at small $Am$, $\beta_0$ must be
sufficiently large $\beta_0\gtrsim100/Am^2$, as can be obtained from equation
(\ref{eq:disp_m}). At a given $Am$, since the stress $\alpha$ monotonically
increases with the net vertical flux, there exist a maximum stress, corresponding
to the largest allowed net flux (smallest allowed value of $\beta_0$). This maximum
value of $\alpha$ as a function of $Am$ is illustrated in the dashed line by connecting
results from runs Z7e, Z6w, Z5 and Z3s. We see that the maximum $\alpha$ drops by
a factor of about $40$ from the ideal MHD case to $Am=1$, and another factor of about
$60$ as $Am$ decreases to $0.1$. By extrapolating this trend, we expect the MRI
turbulence can be self-sustained for arbitrarily small value of $Am$, as long as the
background magnetic field is sufficiently weak. Nevertheless, the turbulence would
seem to be too weak ($\alpha<10^{-4}$) to produce significant amount of angular
momentum transport as required by most astrophysical disks.

\subsubsection[]{The Effect of $\nu$}\label{sssec:nu}

The parameter $\nu$ reflects the sensitivity of how the AD coefficient depends on
gas density (see Equation (\ref{eq:rhoi})). Most of our simulations are run with fixed
value of $\nu=0.5$, while $\nu$ can in principal span a range from $0$ to $1$. The
significance about the effect of $\nu$ largely depends on the level of density
fluctuation in the MRI turbulence. In Table \ref{tab:turbBz}, we list the rms density
fluctuation relative to the background gas density from all vertical net flux runs (see
column $\langle\delta\rho\rangle/\rho_0$). The rms density fluctuation in the ideal
MHD case (run Z1) is relatively large, up to $0.3$, and the largest and smallest
densities reach about $0.2$ and $4$ times the background density. Since AD
reduces the saturation level of the MRI turbulence, the density fluctuations become
smaller as $Am$ decreases. This fact undermines the importance of $\nu$: when
the effect of $\nu$ may be important (large density fluctuations), AD only plays an
insignificant role in the MRI turbulence (large $Am$); when AD strongly affect the
MRI turbulence (small $Am$), the density fluctuation becomes much smaller and
$\nu$ is much less likely to be important. This above implies that variations in the
value of $\nu$ should not have a major impact,
and in particular, the critical value of $Am$ below which MRI is suppressed is
unlikely to be altered by different choices of $\nu$.

To confirm our expectations, we perform two additional runs with the same initial
conditions as run Z5 ($Am=1$, $\beta_0=400$), but set $\nu$ to be $0$
and $1$ respectively. These two runs are named Z5a and Z5b. We see from
Table \ref{tab:turbBz} that the turbulence properties from these two runs are
essentially identical to those in run Z5. Even though our time averages are taken
over relatively short periods, the deviations are generally within $10\%$. This is
understandable since the density fluctuations in these runs are as small as $0.07$.
It appears certain that the value of $\nu$ only plays a very minor role in the MRI
turbulence in the strong coupling limit.

\begin{table*}
\caption{Time and volume averaged quantities in net vertical flux
simulations.}\label{tab:turbBz}
\begin{center}
\begin{tabular}{cccccccccccc}\hline\hline
 Run &$E_{k,x}$ & $E_{k,y}$ & $E_{k,z}$ & $E_{k}$ & $E_{M,x}$  & $E_{M,y}$
 & $E_{M,z}$ & $\langle\delta\rho\rangle/\rho_0$ & $\alpha_{\rm Re}$
 & $\alpha_{\rm Max}$ & $\alpha$ \\\hline
 Z1 & $7.0\times10^{-2}$ & $0.10$                     & $2.9\times10^{-2}$ & $0.20$
       & $8.4\times10^{-2}$ & $0.20$                     & $3.4\times10^{-2}$ & $0.36$
       & $5.4\times10^{-2}$ & $0.17$                     & 0.22 \\
      
 Z2 & $4.4\times10^{-2}$ & $6.5\times10^{-2}$ & $1.9\times10^{-2}$ & $0.13$
       & $4.5\times10^{-2}$ & $8.7\times10^{-2}$ & $2.1\times10^{-2}$ & $0.31$
       & $3.3\times10^{-2}$ & $7.4\times10^{-2}$ & 0.11 \\

 Z3 & $2.8\times10^{-2}$ & $2.7\times10^{-2}$ & $1.4\times10^{-2}$ & $6.9\times10^{-2}$
       & $1.4\times10^{-2}$ & $3.3\times10^{-2}$ & $9.9\times10^{-3}$ & $0.20$
       & $1.6\times10^{-2}$ & $2.9\times10^{-2}$ & $4.5\times10^{-2}$ \\
       
 Z4 & $1.8\times10^{-2}$ & $1.3\times10^{-2}$ & $8.6\times10^{-3}$ & $4.0\times10^{-2}$
       & $6.5\times10^{-3}$ & $1.8\times10^{-2}$ & $6.3\times10^{-3}$ & $0.14$
       & $9.0\times10^{-3}$ & $1.5\times10^{-2}$ & $2.4\times10^{-2}$ \\
       
 Z5 & $5.4\times10^{-3}$ & $1.9\times10^{-3}$ & $3.1\times10^{-3}$ & $1.0\times10^{-2}$
       & $1.2\times10^{-3}$ & $5.1\times10^{-3}$ & $3.3\times10^{-3}$ & $0.070$
       & $2.4\times10^{-3}$ & $4.1\times10^{-3}$ & $6.5\times10^{-3}$ \\
       
 Z3s & $3.5\times10^{-2}$ & $3.3\times10^{-2}$ & $9.6\times10^{-3}$ & $7.8\times10^{-2}$
       & $3.4\times10^{-2}$ & $5.0\times10^{-2}$ & $2.7\times10^{-2}$ & $0.18$
       & $2.9\times10^{-2}$ & $4.0\times10^{-2}$ & $6.9\times10^{-2}$ \\
       
 Z5s & $3.7\times10^{-3}$ & $1.8\times10^{-3}$ & $3.5\times10^{-3}$ & $9.0\times10^{-3}$
       & $1.2\times10^{-3}$ & $1.2\times10^{-3}$ & $1.0\times10^{-2}$ & $0.063$
       & $2.4\times10^{-3}$ & $2.2\times10^{-3}$ & $4.6\times10^{-3}$ \\
        
 Z3w & $1.0\times10^{-2}$ & $7.3\times10^{-3}$ & $4.6\times10^{-3}$ & $2.2\times10^{-2}$
       & $4.4\times10^{-3}$ & $2.0\times10^{-2}$ & $3.6\times10^{-3}$ & $0.091$
       & $5.1\times10^{-3}$ & $1.2\times10^{-2}$ & $1.7\times10^{-2}$ \\
 
 Z5w & $2.5\times10^{-3}$ & $1.1\times10^{-3}$ & $9.6\times10^{-4}$ & $4.6\times10^{-3}$
       & $3.6\times10^{-4}$ & $3.5\times10^{-3}$ & $1.1\times10^{-3}$ & $0.054$
       & $9.5\times10^{-4}$ & $1.5\times10^{-3}$ & $2.5\times10^{-3}$ \\
       
 Z6w & $9.3\times10^{-4}$ & $2.0\times10^{-4}$ & $5.8\times10^{-4}$ & $1.7\times10^{-3}$
       & $5.3\times10^{-5}$ & $7.7\times10^{-4}$ & $7.2\times10^{-4}$ & $0.037$
       & $2.9\times10^{-4}$ & $3.3\times10^{-4}$ & $6.2\times10^{-4}$ \\
       
 Z6e & $7.0\times10^{-4}$ & $1.5\times10^{-4}$ & $7.7\times10^{-5}$ & $9.2\times10^{-4}$
       & $2.1\times10^{-5}$ & $4.7\times10^{-4}$ & $1.5\times10^{-4}$ & $0.036$
       & $2.1\times10^{-4}$ & $1.2\times10^{-4}$ & $3.3\times10^{-4}$ \\
       
 Z7e & $2.8\times10^{-4}$ & $5.3\times10^{-5}$ & $3.4\times10^{-5}$ & $3.7\times10^{-3}$
       & $4.3\times10^{-6}$ & $1.3\times10^{-4}$ & $1.2\times10^{-4}$ & $0.023$
       & $6.6\times10^{-5}$ & $3.1\times10^{-5}$ & $9.7\times10^{-5}$ \\
       
 Z5a & $5.4\times10^{-3}$ & $2.0\times10^{-3}$ & $3.3\times10^{-3}$ & $1.1\times10^{-2}$
       & $1.2\times10^{-3}$ & $5.2\times10^{-3}$ & $3.3\times10^{-3}$ & $0.070$
       & $2.4\times10^{-3}$ & $4.1\times10^{-3}$ & $6.4\times10^{-3}$ \\
       
 Z5b & $5.0\times10^{-3}$ & $1.8\times10^{-3}$ & $3.3\times10^{-3}$ & $1.0\times10^{-2}$
       & $1.4\times10^{-3}$ & $4.9\times10^{-3}$ & $3.3\times10^{-3}$ & $0.068$
       & $2.2\times10^{-3}$ & $3.8\times10^{-3}$ & $6.0\times10^{-3}$ \\
\hline\hline
\end{tabular}
\end{center}
\end{table*}

\subsubsection[]{Properties of the MRI turbulence with AD}\label{sssec:turbprop}

Besides the general properties of the MRI turbulence listed in Table \ref{tab:turbBz},
we study two other aspects of the MRI turbulence with AD.

First, we study the power spectrum density (PSD) of magnetic and kinetic energies
by Fourier analysis. The Fourier analysis in the shearing periodic system is
performed by the remapping technique before and after Fourier transformation, as
described in Section 2.4 of \citet{HGB95}. Although the PSD is anisotropic in
$k-$space, it would be beneficial to plot the PSD in one dimensional form by some
averaging procedure. Following \citet{Davis_etal10}, we compute shell-integrated
power spectrum of the magnetic field $B_k^2\equiv4\pi k^2|\wt{\mb B}(k)|^2$, where
$|\wt{\mb B}(k)|^2$ denotes the average of $|\wt{\mb B}({\mb k})|^2$ over shells of
constant $k=|{\mb k}|$, and
$\wt{\mb B}({\mb k})=\int {\mb B}({\mb x})e^{-i{\mb k}\cdot{\mb x}}d^3{\mb x}/V$ is
the Fourier transform of ${\mb B}({\mb x})$. Here $V$ is the volume of the
simulation box. The Fourier transformation is of course discrete, but for notational
convenience we write the formulas in continuous form. According to Parseval's
theorem, we have
\begin{equation}
\begin{split}
\frac{1}{V}\int_V |{\mb B}({\mb x})|^2d^3{\mb x}&=\int|\wt{\mb B}({\mb k})|^2
\frac{d^3{\mb k}}{(2\pi)^3}\\
&=\frac{1}{(2\pi)^3}\int 4\pi k^2|\wt{\mb B}(k)|^2dk\ .
\end{split}
\end{equation}
Dividing by a factor of $8\pi$ we obtain the PSD for the magnetic energy density
$M_k=B_k^2/8\pi$. Similarly, one can obtain the PSD for the kinetic energy
$K_k$. 

\begin{figure}
    \centering
    \includegraphics[width=92.5mm]{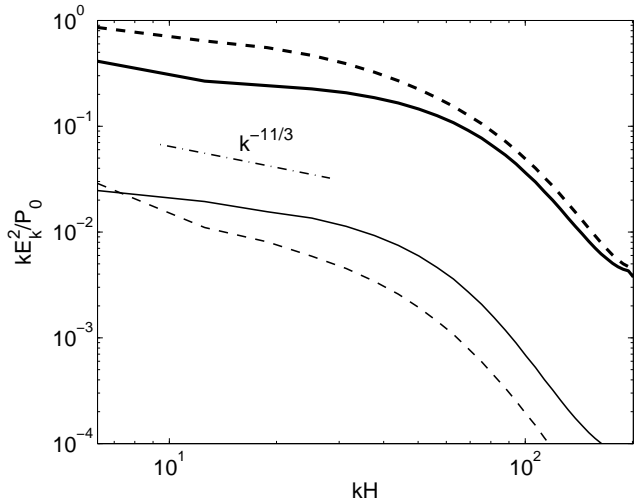}
  \caption{The power spectrum density of the kinetic (solid) and magnetic (dashed)
  energy densities, for two vertical net flux simulations with $Am=1000$ (run Z1, bold)
  and $Am=1$ (run Z5, thin). Plotted are the shell integrated spectrum, represented by
  $E_k^2=4\pi k^2|{\wt{\mb E}}(k)|^2$ (where $E$ denotes kinetic or magnetic energy
  density), normalized to background pressure $P_0=\rho_0c_s^2$. The area enclosed
  by each curve corresponds to the total energy density from turbulent fluctuations.
  }\label{fig:pwr_Bz}
\end{figure}

In Figure \ref{fig:pwr_Bz}, we show the PSDs computed from our runs Z1 and Z5.
These two simulations are representative for the MRI turbulence in the ideal MHD
and AD dominated regimes respectively, and are run for two times longer than many
other simulations (thus giving better statistics). We see that the shape of the PSD
obtained from our simulations are very similar. The PSD roughly follows a power-law
form at small $k$, with the power law index approximately equals to $-11/3$, which is
the index for incompressible Kolmogorov turbulence spectrum.
There appears to be a spectral break at
$kH\approx70$, corresponding to a wavelength of about $0.1H$, and the PSD falls
off rapidly toward smaller scales. The turbulent power in the $Am=1$ case is about
$20$ times smaller than that in the ideal MHD case. Magnetic energy fluctuations
dominate kinetic energy fluctuations in the ideal MHD case, while in the AD
dominated regime, more turbulence power resides in the kinetic energy. Moreover,
we have also checked the contour plot of vertically integrated PSD (not shown) and
found that the turbulence becomes more anisotropic in the AD dominated regime: the
turbulent power is more elongated in $k_x$ than in $k_y$.

\begin{figure}
    \centering
    \includegraphics[width=92.5mm]{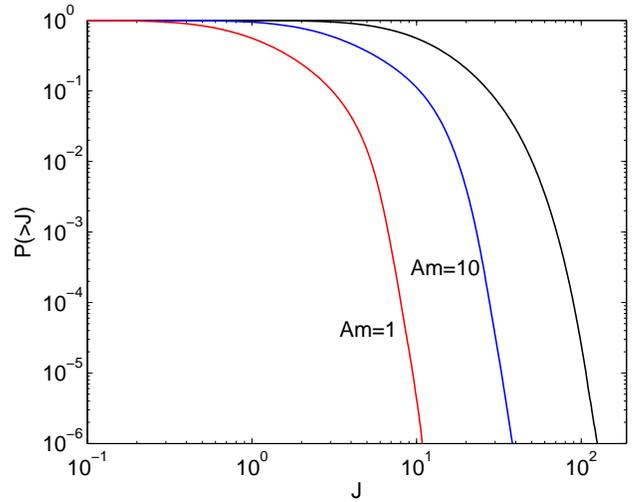}
  \caption{Cumulative probability distribution of current density $J$ for our simulations
  Z1 ($Am=1000$, black), Z3 ($Am=10$, blue), Z5 ($Am=1$, red). The current is
  normalized to $\sqrt{4\pi P_0}c/4\pi H$.
  }\label{fig:Jtot}
\end{figure}

\begin{figure}
    \centering
    \includegraphics[width=92.5mm]{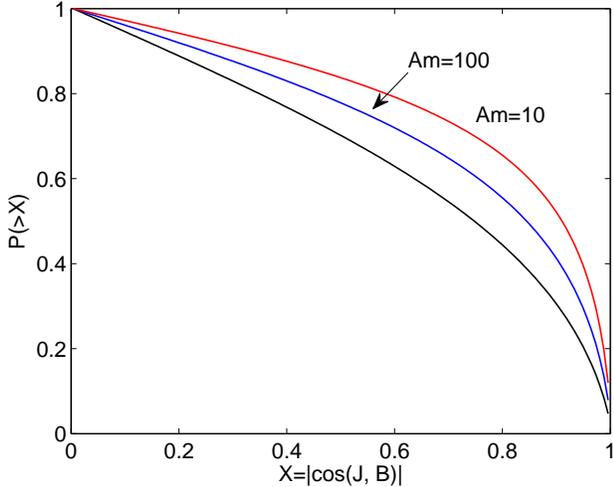}
  \caption{Cumulative probability distribution of $|\cos\alpha|$, where $\alpha$ is the
  angle between ${\mb J}$ and ${\mb B}$. Shown are the results from our runs Z1
  ($Am=1000$, black), Z2 ($Am=100$, blue), Z3 ($Am=10$, red).
  }\label{fig:JB}
\end{figure}

Second, we study the effect of AD on the distribution of current in the MRI turbulence.
It has been shown that in one and two dimensions, sharp current structure can be
developed around magnetic nulls in the presence of AD \citep{BrandenburgZweibel94}.
To examine whether the same effect is present in the MRI turbulence, we show in
Figure \ref{fig:Jtot} the cumulative probability distribution of the current density
$J=|{\mb J}|$ in our simulation runs Z1, Z3 and Z5. If sharp current structure were to
form, one would expect to see extended tails in the probability distribution. However,
we see that as $Am$ decreases, the probability distribution shifts leftward since
turbulence becomes weaker, but its shape remains largely unchanged. We also note
that the current sharpening phenomenon is not observed in simulations by
\citep{Brandenburg_etal95} either. It is likely the sharpening of current by AD is
overwhelmed in 3D MHD turbulence.

AD has also been shown to tend to reduce the current component that is
perpendicular to the magnetic field \citep{Brandenburg_etal95} and make the magnetic
configuration more force-free. To examine this effect, we show the cumulative
probability distribution of $|\cos\alpha|$ from our simulation runs Z1, Z2 and Z3 in
Figure \ref{fig:JB}, where $\alpha$ is the angle between the current and the magnetic
field. The cumulative distribution functions from our runs Z4 and Z5 are almost identical
to that from run Z3, where $Am=10$. We confirm that AD makes the distribution more
concentrated toward $|\cos\alpha|=1$ (i.e., ${\mb J}\parallel{\mb B}$).\footnote{We also
report a difference in our results from \citet{Brandenburg_etal95}. In their simulations,
the distribution function peaks at $|\cos\alpha|=0$ in ideal the MHD case, and AD
concentrate the current toward $|\cos\alpha|=1$. In our simulations, we find that the
distribution function is already concentrated at $|\cos\alpha|=1$ under ideal MHD, and
AD simply makes it more concentrated.} Nevertheless, since the distribution of
$|\cos\alpha|$ is already peaked at 1 in the essentially ideal MHD run (Z1), the effect
of AD does not modify the distribution of current orientation substantially.

\subsection[]{Net toroidal flux simulations}\label{ssec:netBy}

In the second group of simulations, the initial field configuration is chosen to be
uniform along the azimuthal direction $\hat{y}$, with strength characterized by
$\beta_0=2P_0/B_0^2$, where $B_0$ is the initial field strength. Following
\citet{SimonHawley09},  we fix the box size to be $H\times4H\times H$ in the radial,
azimuthal and vertical dimensions for all simulation runs in this group\footnote{We have
also performed our run series Y1 to Y6 using a larger box $4H\times4H\times H$ and
found that the turbulence properties are very similar.}. We choose the fiducial resolution
to be $64$ cells per $H$ in the radial and vertical direction, and $32$ cells per $H$ in
the azimuthal direction. All our net toroidal flux simulations are listed in Table
\ref{tab:netBy}, including one set of fiducial simulations with $\beta_0=100$, one set of
higher resolution simulations, and one set of weak field simulations with $\beta_0=400$.
Unlike net vertical flux, the net toroidal flux is not precisely conserved in our shearing
box simulations. As discussed in \citep{SimonHawley09}, ensuring strict conservation of
toroidal flux numerically is more complex, and is also less important than conserving net
vertical flux because the saturation level of the MRI turbulence is not very sensitive to the
toroidal flux. Throughout all our simulations in this group, we find that the deviation of net
toroidal flux from the initial value is generally less than $2\%$.  

\begin{table}
\caption{Net toroidal flux simulations.}\label{tab:netBy}
\begin{center}
\begin{tabular}{cccccccc}\hline\hline
 Run & $Am$ & $\beta_0$ & Resolution & Restart &Turbulence \\\hline
Y1 & 1000 & 100 & $64\times128\times64$ & No & Yes \\
Y2 & 100   & 100 & $64\times128\times64$ & No & Yes \\
Y3 & 10     & 100 & $64\times128\times64$ & Y1 & Yes \\
Y4 & 3.33  & 100 & $64\times128\times64$ & Y1 & Yes \\
Y5 & 1       & 100 & $64\times128\times64$ & Y1 & No \\
Y6 & 0.33  & 100 & $64\times128\times64$ & Y1 & No \\
Y7 & 0.1     & 400 & $64\times128\times64$ & Y1 & No \\
Y1w & 1000 & 100 & $64\times128\times64$ & No & Yes \\
Y2w & 100 &  100 & $64\times128\times64$ & No & Yes \\
Y3w & 10  & 400 & $64\times128\times64$ & Y1w & Yes \\
Y4w & 3.33 & 400 & $64\times128\times64$ & Y1w & Yes \\
Y5w & 1     & 400 & $64\times128\times64$ & Y1w & No \\
Y6w & 0.33 & 400 & $64\times128\times64$ & Y1w & No \\
Y7w & 0.1   & 400 & $64\times128\times64$ & Y1w & No \\
Y1h & 1000 & 100 & $128\times256\times128$ & No & Yes \\
Y3h & 10 & 100 & $128\times256\times128$ & Y1h & Yes \\
Y5h & 1 & 100 & $128\times256\times128$ & Y1h & No \\
Y6h & 0.33 & 100 & $128\times256\times128$ & Y1h & No \\
\hline\hline
\end{tabular}
\end{center}
Box size is fixed at $H\times4H\times H$, $\nu$ is fixed at $0.5$.
All simulations are run for $24$ orbits (150$\Omega^{-1}$).
\end{table}

The linear stability of Keplerian disks in the presence of pure toroidal field is more
complex than that for the vertical field case. It requires consideration of
non-axisymmetric perturbations \citep{BalbusHawley92}, and involves the
time-dependent amplification of wave modes as the radial wave number swings
from leading to trailing. In ideal MHD, pure toroidal MRI favors high $k_z$ wave
numbers, and requires relatively large numerical resolution. In the case of Ohmic
resistivity, swing amplification of modes is suppressed when the diffusion time of
the mode is comparable to the orbital frequency \citep{PapTerquem97}. A linear
stability with non-axisymmetric perturbations in AD dominated regime has yet to be
performed. Nevertheless, one might expect that a similar argument holds for AD,
with $Am\sim1$ as the boundary for stability. 

\subsubsection[]{A Fiducial Set of Runs}\label{sssec:fidBy}

We fix $\beta_0=100$ and run 7 fiducial simulations with different $Am$ values,
named from Y1 with $Am=1000$ to Y7 with $Am=0.1$ (see Table \ref{tab:netBy})
similar to the case of net vertical flux runs. Initial growth of the MRI from pure
toroidal field is more difficult and is only achieved when $Am$ is greater than $10$.
We initialize the rest of the simulations from the turbulent state at the end of run Y1.

\begin{figure}
    \centering
    \includegraphics[width=92.5mm]{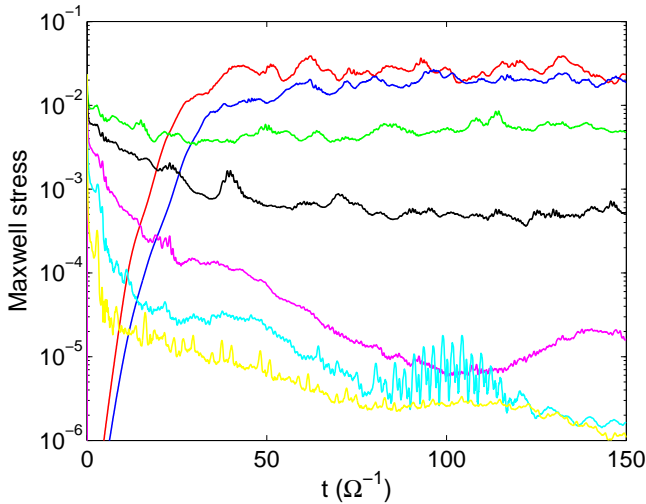}
  \caption{The evolution of Maxwell stress in our fiducial set of toroidal net-flux
  runs (from top to bottom: Y1, Y2, ..., Y7) normalized to $c_sH$. For all runs after $Y2$,
  the simulations start from the end of the Y1 run.}\label{fig:alphByhist}
\end{figure}

Figure \ref{fig:alphByhist} illustrates the time evolution of the Maxwell stress from
this fiducial set of runs. The time and volume averaged quantities from these runs
are listed in Table \ref{tab:turbBy}. We find that at a given value of $Am$, the
saturation level of the MRI with net toroidal flux is much lower than the net vertical
flux case. The turbulent energy density and total stress in run Y1 (essencially
ideal MHD) is a few times $10^{-2}$, about an order of magnitude less than run Z1.
As $Am$ drops below $10$, the saturation level of the MRI turbulence falls off
very rapidly. At $Am=3$ (run Y4), the total stress falls below $10^{-3}$. At $Am=1$,
although a total stress is maintained at a level of $10^{-5}$, we do not observe
any signature of the MRI turbulence by examining the structure of the velocity field,
which is essentially laminar (see further discussion in Section \ref{sssec:criterion}).
Since the simulation is initialized from a turbulent state, the low level of stress and
kinetic energy are mostly due to the eigen-modes of the shearing box excited from
the initial turbulence, and that are not damped due to the low dissipation in the Athena
code. Unlike the case for net vertical flux, there appears to exist a critical value of
$Am$ below which MRI turbulence with net toroidal flux is not self-sustained. This
critical value of $Am$ is about $3$. This fact will be further discussed shortly in
Section \ref{sssec:strBy}.


HS also performed a number of net toroidal flux two-fluid simulations with
$Am\approx1$ and $Am\approx100$, and in both cases turbulence is self-sustained
with total stress $\alpha$ on the order of $10^{-4}$ to $10^{-3}$. Again, these results
are no longer valid in the strong-coupling regime and are not directly comparable to
our results (see discussion in Section \ref{sssec:fidBz}).

We see from Table \ref{tab:turbBy} that the density fluctuations in the net toroidal
flux simulations are generally smaller than those in the net vertical flux case.
Therefore, following the discussion in Section \ref{sssec:nu}, we expect the effect
of $\nu$ has essentially no impact on the conclusions we have drawn above.

\begin{table*}
\caption{Time and volume averaged quantities in net toroidal flux
simulations.}\label{tab:turbBy}
\begin{center}
\begin{tabular}{cccccccccccc}\hline\hline
 Run & $E_{k,x}$ & $E_{k,y}$ & $E_{k,z}$ & $E_k$ & $E_{M,x}$ & $E_{M,y}$ & $E_{M,z}$
  & $\langle\delta\rho\rangle/\rho_0$ & $\alpha_{\rm Re}$ & $\alpha_{\rm Max}$
 & $\alpha$ \\\hline
 Y1 & $1.2\times10^{-2}$ & $1.1\times10^{-2}$ & $5.2\times10^{-3}$ & $2.8\times10^{-2}$
       & $9.7\times10^{-3}$ & $5.0\times10^{-2}$ & $4.4\times10^{-3}$ & $0.10$
       & $7.4\times10^{-3}$ & $2.6\times10^{-2}$ & $3.4\times10^{-2}$ \\
      
 Y2 & $1.0\times10^{-2}$ & $8.2\times10^{-3}$ & $4.6\times10^{-3}$ & $2.3\times10^{-2}$
       & $8.2\times10^{-3}$ & $4.0\times10^{-2}$ & $4.0\times10^{-3}$ & $0.083$
       & $6.2\times10^{-3}$ & $2.0\times10^{-2}$ & $2.6\times10^{-2}$ \\

 Y3 & $4.6\times10^{-3}$ & $2.4\times10^{-3}$ & $1.8\times10^{-3}$ & $8.8\times10^{-3}$
       & $2.4\times10^{-3}$ & $1.9\times10^{-2}$ & $1.4\times10^{-3}$ & $0.062$
       & $2.5\times10^{-3}$ & $5.7\times10^{-3}$ & $8.2\times10^{-3}$ \\
       
 Y4 & $6.9\times10^{-4}$ & $2.8\times10^{-4}$ & $2.1\times10^{-4}$ & $1.2\times10^{-3}$
       & $3.4\times10^{-4}$ & $1.0\times10^{-2}$ & $2.4\times10^{-4}$ & $0.029$
       & $3.0\times10^{-4}$ & $5.0\times10^{-4}$ & $8.1\times10^{-4}$ \\
       
Y5$^*$
       & $3.5\times10^{-5}$ & $8.2\times10^{-5}$ & $1.8\times10^{-5}$ & $1.4\times10^{-4}$
       & $1.4\times10^{-5}$ & $9.7\times10^{-3}$ & $3.1\times10^{-5}$ & $0.008$
       & $6.6\times10^{-6}$ & $1.2\times10^{-5}$ & $1.9\times10^{-5}$ \\
       
Y1w & $7.3\times10^{-3}$ & $5.6\times10^{-3}$ & $3.1\times10^{-3}$ & $1.6\times10^{-2}$
       & $4.8\times10^{-3}$ & $2.8\times10^{-2}$ & $2.1\times10^{-3}$ & $0.079$
       & $4.5\times10^{-3}$ & $1.5\times10^{-2}$ & $1.9\times10^{-2}$ \\
       
Y2w & $4.9\times10^{-3}$ & $2.8\times10^{-3}$ & $1.8\times10^{-3}$ & $9.6\times10^{-3}$
       & $2.5\times10^{-3}$ & $1.6\times10^{-2}$ & $1.2\times10^{-3}$ & $0.063$
       & $2.7\times10^{-3}$ & $7.7\times10^{-3}$ & $1.0\times10^{-2}$ \\
       
Y3w & $1.7\times10^{-3}$ & $5.8\times10^{-4}$ & $4.7\times10^{-4}$ & $2.7\times10^{-3}$
       & $4.9\times10^{-4}$ & $5.4\times10^{-3}$ & $2.6\times10^{-4}$ & $0.043$
       & $7.2\times10^{-4}$ & $1.5\times10^{-3}$ & $2.2\times10^{-3}$ \\
       
Y4w & $3.3\times10^{-4}$ & $8.1\times10^{-5}$ & $8.1\times10^{-5}$ & $4.9\times10^{-4}$
       & $6.8\times10^{-5}$ & $3.4\times10^{-3}$ & $5.8\times10^{-5}$ & $0.022$
       & $1.0\times10^{-4}$ & $2.2\times10^{-4}$ & $3.2\times10^{-4}$ \\
       
Y5w$^*$
       & $1.2\times10^{-4}$ & $6.4\times10^{-5}$ & $1.4\times10^{-5}$ & $2.0\times10^{-4}$
       & $5.6\times10^{-6}$ & $2.4\times10^{-3}$ & $4.2\times10^{-6}$ & $0.014$
       & $2.4\times10^{-5}$ & $1.1\times10^{-5}$ & $3.5\times10^{-5}$ \\
       
Y1h & $1.4\times10^{-2}$ & $1.7\times10^{-2}$ & $6.7\times10^{-3}$ & $3.7\times10^{-2}$
       & $1.6\times10^{-2}$ & $7.3\times10^{-2}$ & $8.1\times10^{-3}$ & $0.12$
       & $8.4\times10^{-3}$ & $3.8\times10^{-2}$ & $4.7\times10^{-2}$ \\
       
Y3h & $4.5\times10^{-3}$ & $2.9\times10^{-3}$ & $1.9\times10^{-3}$ & $9.3\times10^{-3}$
       & $3.2\times10^{-3}$ & $2.3\times10^{-2}$ & $1.9\times10^{-3}$ & $0.056$
       & $2.5\times10^{-3}$ & $6.8\times10^{-3}$ & $9.3\times10^{-3}$ \\
       
Y5h$^*$
       & $7.4\times10^{-5}$ & $6.3\times10^{-5}$ & $8.7\times10^{-6}$ & $1.5\times10^{-4}$
       & $1.0\times10^{-5}$ & $1.0\times10^{-2}$ & $3.2\times10^{-5}$ & $0.011$
       & $1.5\times10^{-5}$ & $7.5\times10^{-6}$ & $2.2\times10^{-5}$ \\
\hline\hline
\end{tabular}
\end{center}
$^*$: These runs are not turbulent.

\end{table*}

\subsubsection[]{The Effect of Resolution}\label{sssec:convergence}

Relatively high resolution is needed for net toroidal flux MRI simulations in order to
properly capture the amplification of wave modes as they swing from leading to trailing
\citep{SimonHawley09}. In order to justify our results in the previous subsection, we
perform a few of the simulations with doubled resolution. These runs are labeled with
an additional letter ``h" (i.e., high resolution) in Table \ref{tab:netBy}. 

The time and volume averaged quantities from these high resolution runs are shown
in Table \ref{tab:turbBy}. We see that the kinetic and magnetic energy densities from in
the high resolution simulations are generally larger than those in the low resolution
runs, but only by a small factor. In particular, the difference between low and high
resolutions with relatively large AD coefficient is only about $10\%$ (e.g., comparing
runs Y3 and Y3h), which strongly indicates numerical convergence. This is not
surprising since small scale structures can be largely damped by AD thus higher
resolution becomes unnecessary. Run Y5h is also very similar to run Y5, where the
initial turbulence is damped with remnant small velocity and magnetic fluctuations
unlikely to be associated with the MRI turbulence.

\begin{figure}
    \centering
    \subfigure{
    \includegraphics[width=92.5mm]{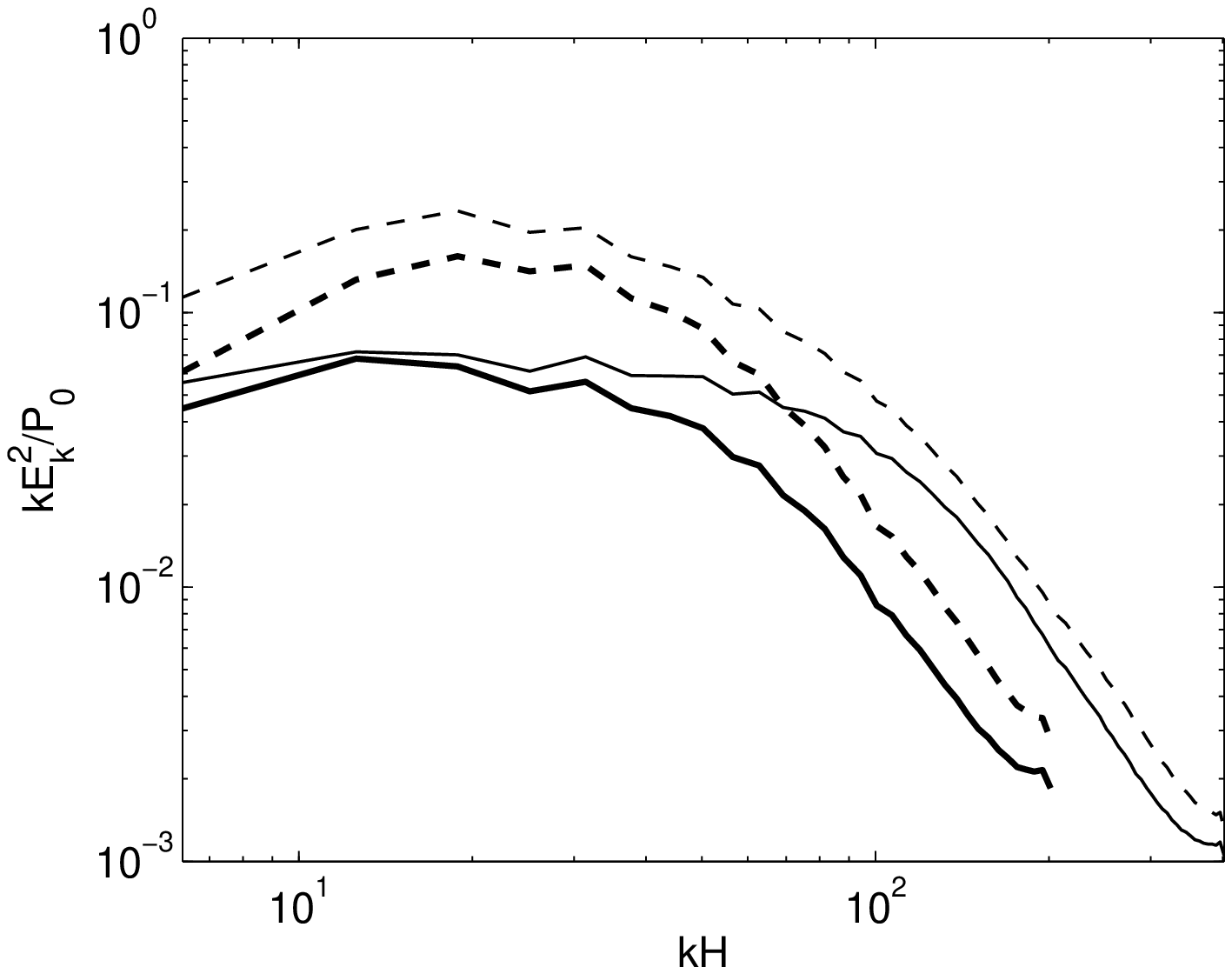}}
    \subfigure{
    \includegraphics[width=92.5mm]{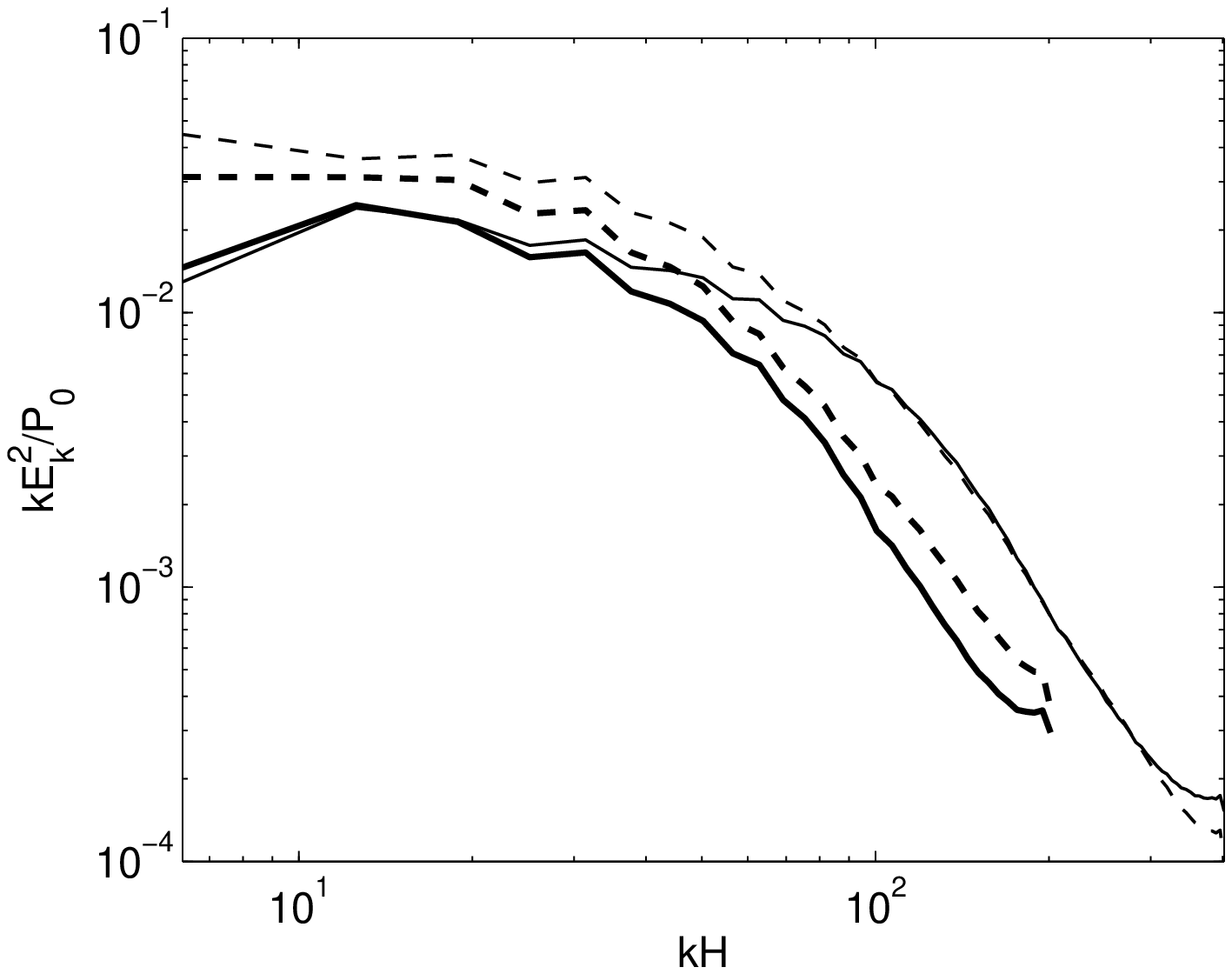}}
  \caption{Similar to Figure \ref{fig:pwr_Bz}, but for net toroidal flux simulations. Shown
  are the shell integrated power spectrum density of the kinetic (solid) and magnetic
  (dashed) energy densities. Upper panel: results from simulations with $Am=1000$
  (essentially ideal MHD). Our fiducial low-resolution results (run Y1) are shown in
  bold curves, while high-resolution results (run Y1h) are shown in thin lines. Lower
  panel: results from simulations with $Am=10$, with low-resolution (run Y3) and
  high-resolution (Y3h) results shown in bold and thin lines. The uncertainty is about
  $10\%$.}\label{fig:pwr_By}
\end{figure}

The inferences above are further justified by looking at the power spectrum of magnetic
and kinetic energies. Following the same procedure described in Section
\ref{sssec:turbprop}, we show in Figure \ref{fig:pwr_By} the shell integrated PSDs for
runs Y1, Y1h and Y3, Y3h. The spectral shapes from the low resolution simulations
appear to have a spectral peak at relatively large scales of about $0.5H$, while at higher
resolution, a power law spectrum at intermediate scales from approximately $0.1H$ to
$0.5H$ analogous to an inertia range appears to have developed. This observation
may suggest high numerical resolution of at least $128$ cells per $H$ is needed for the
toroidal field MRI simulations in order to resolve the inertia range in the turbulent
spectrum, although smaller resolution of $64$ cells per $H$ appears to be sufficient
for turbulence properties to converge.
The shape of the PSDs  at small $k$ look very different from the PSDs in the net vertical
flux simulations, indicating different energy injection mechanism, while at large $k$, the
PSDs from both cases fall off in a similar manner, indicating similar dissipation mechanism.

\subsubsection[]{The Effect of toroidal Field Strength}\label{sssec:strBy}

We have also performed the same set of net toroidal flux simulations with the toroidal
magnetic field strength lowered by one half, $\beta_0=400$, labeled with an additional
letter ``w" (i.e., weak field) in Table \ref{tab:netBy}. General properties from the saturated
state of these runs are also shown in Table \ref{tab:turbBy}. We see that at the same
value of $Am$, the kinetic and magnetic energy density from the weak field simulations
are smaller than those in our fiducial runs by a factor of $2$-$3$. By inspection of the
velocity field as well as the distribution of current density, we find that sustained
turbulence is supported in run Y4w but not in run Y5w. Note that the time and volume
averaged kinetic energy density from run Y4w is $4.9\times10^{-4}$, which is slightly
below our limit of $10^{-3}$, but the total stress $\alpha=3.2\times10^{-4}$ is
reasonably large and is unlikely to be caused by inertia waves in the simulation box.
Further discussion will be given in the next subsection \ref{sssec:criterion}.

\begin{figure}
    \centering
    \includegraphics[width=92.5mm]{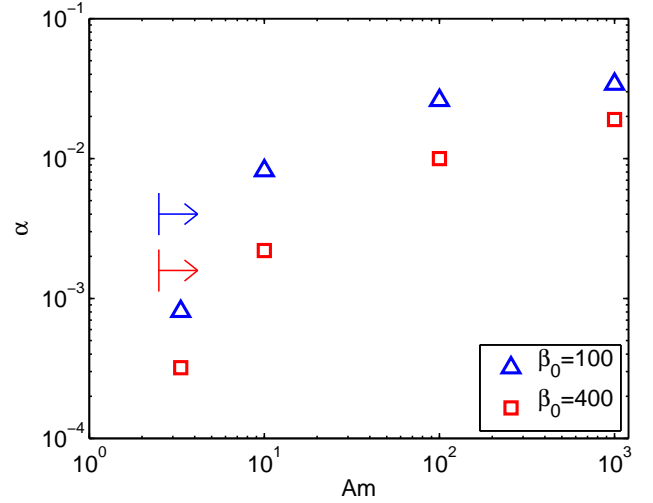}
  \caption{The time and volume averaged total stress $\alpha$ from our net toroidal
  flux simulations. Simulations with different net toroidal flux, characterized by the
  plasma $\beta_0$, are labeled by different symbols and colors. The arrows for each
  $\beta_0$ (as represented in red and blue) indicate the range of $Am$ where MRI
  turbulence can be sustained.}\label{fig:alphByall}
\end{figure}

Together with the results from Section \ref{sssec:fidBy}, we summarize the results
from net toroidal flux simulations in Figure \ref{fig:alphByall}. We see that for both
values of $\beta_0$ (100 and 400), we do not observe any sustained MRI turbulence
for $Am<3$. Although we have not explored weaker net toroidal flux,
based on the trend we see from the Figure, even if MRI can be self-sustained at
$Am\lesssim1$ with weaker net toroidal flux, the resulting total stress $\alpha$ is
unlikely to be above $10^{-4}$. This is in stark contrast with the net vertical flux
simulations, and indicates that pure toroidal field geometry is more stable in the AD
dominated regime.

\subsubsection[]{Criterion for Sustained Turbulence}\label{sssec:criterion}

A key objective of this paper is to study when MRI turbulence can be self-sustained in the
presence of AD in the strong coupling limit. For the net vertical flux simulations, the criterion
is relatively clear: turbulence can be sustained as long as the most MRI unstable mode fits
into the simulation box. 
The situation is less clear for net toroidal flux simulations, and it remains to be seen
whether the latter can be characterized as self-sustained turbulence.

\begin{figure}
    \centering
    \includegraphics[width=92.5mm]{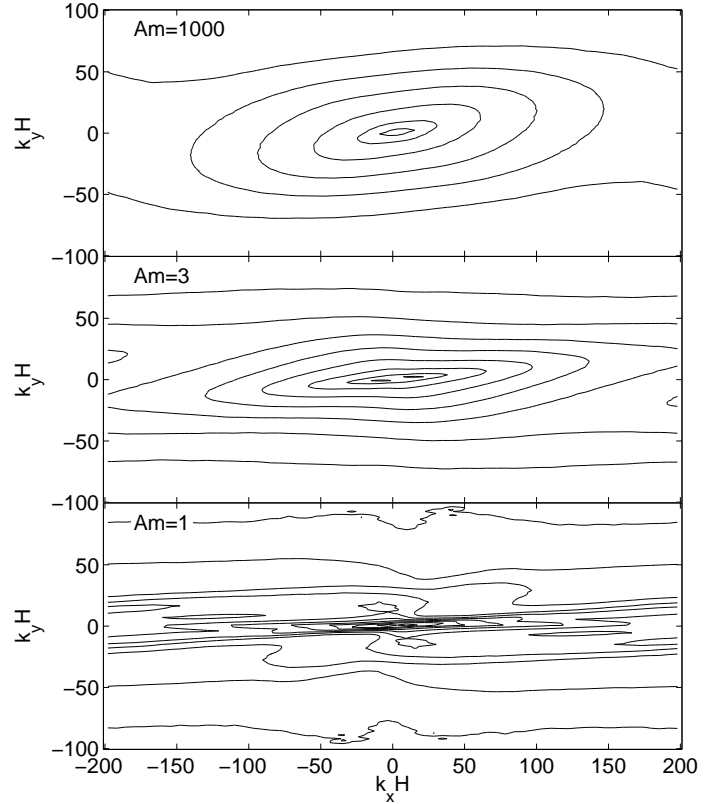}
  \caption{Contour plot of the vertically integrated power spectrum density of kinetic
  energy from our net toroidal flux simulation runs Y1 ($Am=1000$), Y4 ($Am=3$) and
  Y5 ($Am=1)$. Neighboring contours are separated by a factor of 10 in the power
  spectrum density.}\label{fig:criterion}
\end{figure}

We take run Y5 as an illustrative example in this subsection, but the
analysis also applies to other marginal runs including Z6, Y5w, Y5h. In run Y5, after the
initial turbulence has damped, the flow is largely laminar, except in a few very localized
regions where some narrow but azimuthally elongated current structure is present and
evolves very slowly with time. These features can be better demonstrated by computing
the vertically integrated Fourier power spectrum of magnetic and kinetic energies.
Following the same procedure as described in Section \ref{sssec:turbprop} but
integrating the full three-dimensional PSD over $k_z$, we show the contour plot of the
kinetic energy PSD in the $k_x-k_y$ plane from our runs Y1, Y4 and Y5 in Figure
\ref{fig:criterion}. We see that in ideal MHD (Y1), the vertically integrated PSD has
elliptic contours elongated and tilted toward the $k_x$ axis. The contours are distorted
and more elongated when AD is added (Y4). However, in run Y5, we see that the overall
shape of the PSD contours are extremely elongated in the $k_x$ direction. The original
elliptic contours are almost destroyed, with irregular fragments distributed around the
center. These irregular features in the vertically integrated PSD strongly indicate that
the system is not in a turbulent state. We have also found that such irregular features
are also present in runs Z6, Y5w and Y5h, but not present in runs Z5, Y4w and Y3h.

In sum, we conclude that non-zero kinetic energy density and total stress do not prove
the existence of the MRI turbulence in shearing box simulations. Transition from
self-sustained turbulence to non-turbulent state can be judged by looking at the
vertically integrated PSD of kinetic and magnetic energies.

\subsection[]{Simulations with both vertical and toroidal fluxes}\label{ssec:ByBz}

Motivated by the linear stability analysis of the MRI with AD by \citet{KunzBalbus04} and
\citet{Desch04}, we present simulations that include both vertical and toroidal fluxes.
These authors found that in the presence of both vertical and toroidal field, unstable modes
exists for any values of $Am$, and the fastest growth rate is non-vanishing even as
$Am\rightarrow0^+$. When $Am\lesssim1$, the wave number of the most unstable mode
has a substantial non-zero radial component. In this subsection, we explore whether
this behavior in the linear regime affects the non-linear evolution of the MRI with AD.

We use the vertical plasma $\beta$: $\beta_{z0}=8\pi P_0/B_z^2$ to specify the net
vertical magnetic flux. The net toroidal flux is specified by its ratio to the net vertical
field $B_\phi/B_z$. We consider two sets of simulations, with $B_\phi/B_z=4$ and
$B_\phi/B_z=1.25$ and labeled by letters ``M" and ``N" respectively. Parameters of these
runs are given in Table \ref{tab:mixB}. Similar to the previous subsections, we scan
the parameter $Am$ from $1000$ down to $0.1$. We use the dispersion relation
(31) - (35) in \citet{KunzBalbus04} to find the wavenumber of the most unstable mode. We
find that for $Am\gtrsim3$, the most unstable wavenumber is essentially purely vertical,
while for $Am\lesssim1$, the most unstable wavenumber is oblique with $k_z\sim-k_r$.
Correspondingly, we choose the box size to be $H\times4H\times H$ for runs with
$Am\geq3$ and $4H\times4H\times H$ for $Am\leq1$. For simulations with $Am\leq1$, we
expect the fastest growth of the MRI to occur in the diagonal direction of the $x-z$ plane.

By default, we fix $\beta_{z0}=1600$ in both sets of simulations. However, as $Am$ falls
below $1$, we find that the fastest growing mode will no longer fit into our simulation
box. Based on the results from Section \ref{ssec:netBz}, MRI turbulence would not be
sustained in this case. Therefore, we increase $\beta_{z0}$ to the value
such that the most unstable mode just fit into the box. In the case of $B_\phi/B_z=4$,
$\beta_{z0}$ is increased to $8000$ and $3\times10^4$ for $Am=0.33$ and $Am=0.1$
respectively. For $B_\phi/B_z=1.25$, we increase $\beta_{z0}$ to $8000$ at $Am=0.1$.
Consequently, for all simulations in this subsection, MRI grows from the initial seed
perturbations, and generates sustained turbulence after saturation. Below we discuss the
properties of the MRI turbulence in the two sets of simulations.

\begin{table}
\caption{Simulations with both net vertical and toroidal fluxes.}\label{tab:mixB}
\begin{center}
\begin{tabular}{cccccccccc}\hline\hline
 Run & $Am$ & Box size & $\beta_{z0}$$^1$ & $B_\phi/B_z$ & Orbits \\\hline
M1 & 1000 & $H\times4H\times H$ & 1600 & 4 & 48 \\
M2 & 100   & $H\times4H\times H$ & 1600 & 4 & 48 \\
M3 & 10     & $H\times4H\times H$ & 1600 & 4 & 36 \\
M4 & 3.33  & $H\times4H\times H$ & 1600 & 4 & 36 \\
M5 & 1       & $4H\times4H\times H$ & 1600 & 4 & 42 \\
M6 & 0.33   & $4H\times4H\times H$ & 8000 & 4 & 96 \\
M7 & 0.1     & $4H\times4H\times H$ & 30000 & 4 & 96 \\
N1 & 1000 & $H\times4H\times H$ & 1600 & 1.25 & 48 \\
N2 & 100   & $H\times4H\times H$ & 1600 & 1.25 & 48 \\
N3 & 10     & $H\times4H\times H$ & 1600 & 1.25 & 48 \\
N4 & 3.33  & $H\times4H\times H$ & 1600 & 1.25 & 48 \\
N5 & 1       & $4H\times4H\times H$ & 1600 & 1.25 & 46 \\
N6 & 0.33   & $4H\times4H\times H$ & 1600 & 1.25 & 96 \\
N7 & 0.1     & $4H\times4H\times H$ & 8000 & 1.25 & 96 \\
\hline\hline
\end{tabular}
\end{center}
The grid resolution is fixed at $64$ cells per $H$. All simulations are
initiated from seed perturbations, and generate sustained turbulence.

$^1$: plasma $\beta$ from the vertical magnetic field.
\end{table}

\begin{table*}
\caption{Time and volume averaged quantities in simulations with both net vertical
and net toroidal fluxes.}\label{tab:turbByz}
\begin{center}
\begin{tabular}{cccccccccccc}\hline\hline
 Run & $E_{k,x}$ & $E_{k,y}$ & $E_{k,z}$ & $E_k$ & $E_{M,x}$ & $E_{M,y}$ & $E_{M,z}$
  & $\langle\delta\rho\rangle/\rho_0$ & $\alpha_{\rm Re}$ & $\alpha_{\rm Max}$
 & $\alpha$ \\\hline
 M1 & $5.6\times10^{-2}$ & $5.8\times10^{-2}$ & $1.9\times10^{-2}$ & $0.13$
       & $6.7\times10^{-2}$ & $0.30$                    & $2.8\times10^{-2}$ & $0.26$
       & $3.4\times10^{-2}$ & $0.17$                    & $0.21$ \\
      
 M2 & $4.2\times10^{-2}$ & $3.4\times10^{-2}$ & $1.2\times10^{-2}$ & $8.8\times10^{-2}$
       & $4.3\times10^{-2}$ & $0.17$                    & $2.0\times10^{-2}$ & $0.16$
       & $2.4\times10^{-2}$ & $0.10$                    & $0.13$ \\

 M3 & $1.6\times10^{-2}$ & $9.2\times10^{-3}$ & $5.7\times10^{-3}$ & $3.1\times10^{-2}$
       & $1.1\times10^{-2}$ & $5.3\times10^{-2}$ & $6.3\times10^{-3}$ & $0.087$
       & $9.1\times10^{-3}$ & $2.8\times10^{-2}$ & $3.7\times10^{-2}$ \\
       
 M4 & $9.5\times10^{-3}$ & $4.7\times10^{-3}$ & $3.7\times10^{-3}$ & $1.8\times10^{-2}$
       & $3.9\times10^{-3}$ & $2.2\times10^{-2}$ & $3.3\times10^{-3}$ & $0.075$
       & $4.7\times10^{-3}$ & $8.7\times10^{-3}$ & $1.3\times10^{-2}$ \\
       
 M5 & $3.9\times10^{-3}$ & $2.8\times10^{-3}$ & $1.4\times10^{-3}$ & $8.1\times10^{-3}$
       & $1.3\times10^{-3}$ & $1.3\times10^{-2}$ & $1.7\times10^{-3}$ & $0.066$
       & $1.7\times10^{-3}$ & $2.8\times10^{-3}$ & $4.5\times10^{-3}$ \\
       
 M6 & $1.7\times10^{-3}$ & $2.1\times10^{-3}$ & $2.2\times10^{-4}$ & $4.0\times10^{-3}$
       & $1.3\times10^{-4}$ & $2.6\times10^{-3}$ & $3.1\times10^{-4}$ & $0.058$
       & $5.6\times10^{-4}$ & $4.0\times10^{-4}$ & $9.6\times10^{-4}$ \\
       
 M7 & $1.4\times10^{-3}$ & $4.0\times10^{-4}$ & $6.5\times10^{-5}$ & $1.9\times10^{-3}$
       & $3.0\times10^{-5}$ & $7.6\times10^{-4}$ & $7.3\times10^{-5}$ & $0.050$
       & $4.8\times10^{-4}$ & $1.3\times10^{-4}$ & $6.1\times10^{-4}$ \\
       
N1 & $3.8\times10^{-2}$ & $3.9\times10^{-2}$ & $1.4\times10^{-2}$ & $9.1\times10^{-2}$
       & $4.2\times10^{-2}$ & $0.20$                      & $1.8\times10^{-2}$ & $0.20$
       & $2.2\times10^{-2}$ & $0.12$                      & $0.14$ \\
    
N2 & $2.8\times10^{-2}$ & $2.2\times10^{-2}$ & $9.4\times10^{-3}$ & $5.9\times10^{-2}$
       & $2.7\times10^{-2}$ & $0.11$                      & $1.3\times10^{-2}$ & $0.13$
       & $1.6\times10^{-2}$ & $6.8\times10^{-2}$ & $8.3\times10^{-2}$ \\
       
N3 & $1.4\times10^{-2}$ & $6.7\times10^{-3}$ & $4.5\times10^{-3}$ & $2.5\times10^{-2}$
       & $7.4\times10^{-3}$ & $3.6\times10^{-2}$ & $4.4\times10^{-3}$ & $0.080$
       & $6.9\times10^{-3}$ & $2.0\times10^{-2}$ & $2.7\times10^{-2}$ \\
       
N4 & $6.3\times10^{-3}$ & $3.4\times10^{-3}$ & $2.9\times10^{-3}$ & $1.3\times10^{-2}$
       & $1.7\times10^{-3}$ & $9.2\times10^{-3}$ & $2.1\times10^{-3}$ & $0.068$
       & $2.6\times10^{-3}$ & $4.6\times10^{-3}$ & $7.3\times10^{-3}$ \\
       
N5 & $3.2\times10^{-3}$ & $1.8\times10^{-3}$ & $1.3\times10^{-3}$ & $6.3\times10^{-3}$
       & $5.2\times10^{-4}$ & $4.7\times10^{-3}$ & $1.3\times10^{-3}$ & $0.065$
       & $1.2\times10^{-3}$ & $1.8\times10^{-3}$ & $3.0\times10^{-3}$ \\
       
N6 & $1.2\times10^{-3}$ & $1.1\times10^{-3}$ & $9.0\times10^{-4}$ & $3.2\times10^{-3}$
       & $1.2\times10^{-4}$ & $1.8\times10^{-3}$ & $8.7\times10^{-4}$ & $0.057$
       & $3.8\times10^{-4}$ & $4.8\times10^{-4}$ & $8.6\times10^{-4}$ \\
       
N7 & $9.9\times10^{-4}$ & $2.4\times10^{-4}$ & $1.1\times10^{-4}$ & $1.3\times10^{-3}$
       & $2.9\times10^{-5}$ & $5.7\times10^{-4}$ & $1.8\times10^{-4}$ & $0.042$
       & $3.0\times10^{-4}$ & $1.6\times10^{-4}$ & $4.7\times10^{-4}$ \\
\hline\hline
\end{tabular}
\end{center}
\end{table*}

The first group of simulations (with $B_\phi/B_z=4$)
are run for at least $36$ orbits.
The initial growth of the MRI from runs M5 ($Am=1$), M6 ($Am=0.33$) and M7
($Am=0.1$) are slower than the ideal MHD limit due to strong effect of AD: the fastest
growth rates $\sigma_m$ predicted from the linear dispersion relation for these
simulation runs are $0.189\Omega^{-1}$, $0.149\Omega^{-1}$ and $0.136\Omega^{-1}$
respectively. This is to be compared with the case with pure vertical field with the same
$Am$, where the corresponding $\sigma_m$ is $0.428\Omega^{-1}$,
$0.218\Omega^{-1}$ and $0.074\Omega^{-1}$ respectively. The presence of both
vertical and toroidal field gives a smaller growth rate at relatively large $Am$, but
$\sigma_m$ decreases much slower with decreasing $Am$ than the pure vertical
field case. At $Am\lesssim0.1$, a field configuration with both net vertical and
toroidal fluxes becomes more favorable than the pure net vertical field geometry by
having substantially larger $\sigma_m$.

In the second group of runs, we choose $B_\phi/B_z=1.25$,
which generates the fastest grow rate at $Am\leq1$ compared with any other values.
The fastest growth rates $\sigma_m$ for runs N5 ($Am=1$), N6 ($Am=0.33$) and N7
($Am=0.1$) are $0.371\Omega^{-1}$, $0.253\Omega^{-1}$ and $0.206\Omega^{-1}$
respectively, which are nearly two times larger than those for $B_\phi/B_z=4$. 

\begin{figure}
    \centering
    \includegraphics[width=92.5mm]{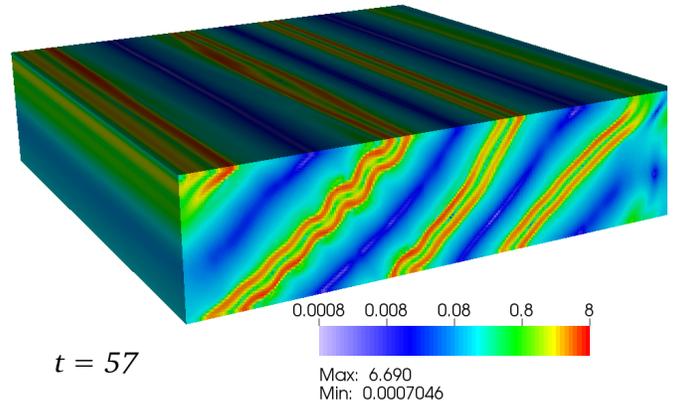}
  \caption{The distribution of current density in run M5 with $Am=1$, $B_\phi/B_z=4$
  at the break down of the ``channel flow" before saturating into turbulence (at time
  $t=57\Omega^{-1}$).}\label{fig:examByz}
\end{figure}

Our simulations with $Am\leq1$ (M5-M7, N5-N7) are particularly interesting because
the fastest growing mode has a non-zero radial wave number $|k_r|$ comparable to
$|k_z|$. During the linear growth stage, we observe axisymmetric structures in the $x-z$
plane similar to channel modes, but tilted toward the diagonal direction. These structures
grow to a large amplitude, and finally break down into turbulence. As an example, we
show in Figure \ref{fig:examByz} the distribution of current density right before the break
down of the channel flow for run M5. In the turbulent state, one can still observe the
emergence of structures elongated in the diagonal direction in the $x-z$ plane from time
to time, which then fragment and inject kinetic energy into the system. For the simulations
with $Am\leq0.33$, these events lead to sporadic increase of kinetic energy and Reynolds
stress on time scales of $10-20$ orbits. Due to such long time variability, we run these
models for longer (to $96$ orbits) and take the time average from about $56$ orbits
($350\Omega^{-1}$) onward.

\begin{figure}
    \centering
    \includegraphics[width=92.5mm]{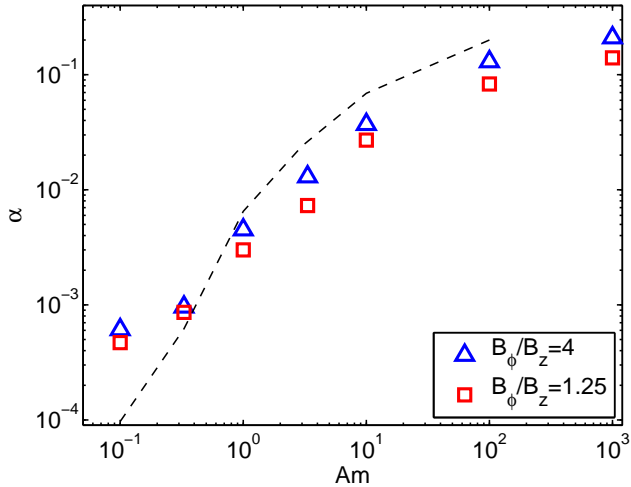}
  \caption{The time and volume averaged total stress $\alpha$ from our simulations
  with both net vertical and net toroidal flux. Shown are results from two groups of
  simulations with different ratio $B_\phi/B_z$, are labeled by different symbols and
  colors. Dashed line represents the maximum value of $\alpha$ obtained from
  pure net vertical flux simulations (Figure \ref{fig:alphBzall}).}\label{fig:alphByzall}
\end{figure}

The time and volume averaged properties of the MRI turbulence in all simulations
with both net vertical and toroidal fluxes are listed in Table \ref{tab:turbByz}. In
Figure \ref{fig:alphByzall} we plot the total turbulent stress $\alpha$ as a function of
$Am$ from the two groups of simulations. We see that at $Am\gtrsim1$, where all
simulations have the same net vertical flux $\beta_{z0}=1600$, runs with
$B_\phi/B_z=4$ generate slightly stronger MRI turbulence than simulations with
$B_\phi/B_z=1.25$. This trend can be extrapolated  down to zero net toroidal flux,
as one can compare with results in runs Z3w and Z5w in Table \ref{tab:turbBz}.
The dependence of turbulent strength on the toroidal flux is relatively
weak, and when the net vertical flux in doubled, as in runs Z1 to Z5, the strength of
the turbulence becomes stronger than our corresponding runs M1 to M5.

For simulations with $Am<1$, we find that the turbulent stress $\alpha$ exceeds the
maximum possible $\alpha$ attainable by the pure net vertical flux simulations. At
$Am=0.1$, the maximum value of $\alpha$ is $6.1\times10^{-4}$, as compared to
about $9.7\times10^{-5}$ from the pure net vertical flux case. This is consistent with
the linear dispersion properties discussed before: the presence of both vertical and
toroidal field raises the maximum growth rate in the $Am<1$ regime. While the values
of $\alpha$ given by the $B_\phi/B_z=4$ group are still larger than those in the
$B_\phi/B_z=1.25$ group, the latter group produces larger Maxwell stress. We note
that for $Am\leq1$, we have chosen the largest possible net flux such that the vertical
extent of the simulation box can fit only one most unstable mode. According to the
discussion in Section \ref{sssec:strBz}, the strength of the MRI turbulence from these
simulations represents the highestpossible level at the given value of $Am$ and the
given magnetic field geometry for our box size, and may also approach the highest
level in real disks.


\section[]{MRI with Ambipolar Diffusion: Diagnostics}

In this section, we combine the results from all our simulations and further discuss the
criteria for whether the MRI can be self-sustained with AD in a more general context.

The MRI acts as a dynamo which amplifies initially weak fields. Although the saturation
mechanism of the MRI is not well understood (but see
\citet{PessahGoodman09,Pessah10}), the magnetic field energy at the saturated state
scales with the net magnetic flux and is generally below equipartition with thermal energy
(\citet{HGB95,Sano_etal98} as well as this paper).
Given the field geometry, there should exist a one-to-one correspondence between the
initial field strength characterized by $\beta_0$ and the final field strength characterized by
$\langle\beta\rangle$ (the space and time averaged gas to magnetic pressure at the
saturated state), with $\langle\beta\rangle<\beta_0$ for weak field due to the MRI dynamo,
and gradually transiting to $\langle\beta\rangle\approx\beta_0$ where the background field
is too strong field to be destabilized.

The quantity $\langle\beta\rangle$ is also very useful for studying non-ideal MHD
effects because the value it controls the relative importance of various non-ideal MHD
effects (Ohmic, Hall and AD, e.g., see \citet{Wardle07,Bai11}). Henceforth, we
shall consider $\langle\beta\rangle$ as a main diagnostic quantity on the MRI turbulence.


\begin{figure}
    \centering
    \includegraphics[width=92.5mm]{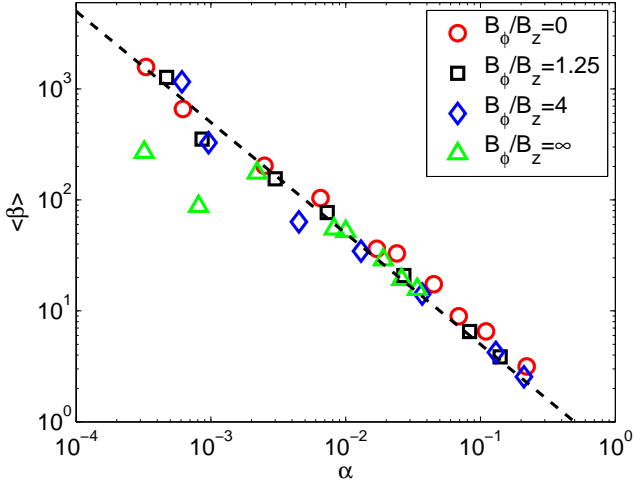}
  \caption{Scattered plot of the total turbulent stress $\alpha$ and the plasma
  $\beta$ at the saturated state of the MRI turbulence from all our simulations.
  Simulations with different field geometries are marked by different symbols and
  colors as indicated in the legend, where $B_\phi/B_z=0$ and $B_\phi/B_z=\infty$
  correspond to pure net vertical and pure net toroidal flux simulations respectively.
  Dashed line shows the fitting curve $\langle\beta\rangle=1/2\alpha$.}
  \label{fig:alphabeta}
\end{figure}

\begin{figure}
    \centering
    \includegraphics[width=92.5mm]{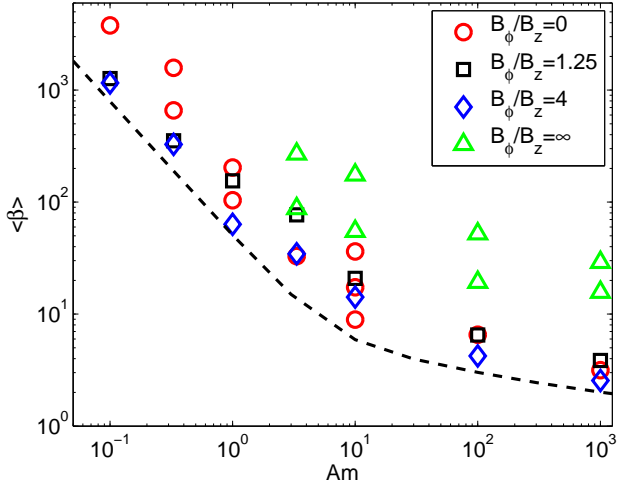}
  \caption{Scattered plot of AD coefficient $Am$ and the plasma $\beta$ at the
  saturated state of the MRI turbulence from all our simulations. Simulations with
  different field geometries are marked by different symbols and colors as indicated
  in the legend. The dashed curve shows the fitting formula (\ref{eq:constrain}) as a
  lower bound of $\langle\beta\rangle\geq\beta_{\rm min}(Am)$.}\label{fig:Ambeta}
\end{figure}

\begin{figure}
    \centering
    \includegraphics[width=92.5mm]{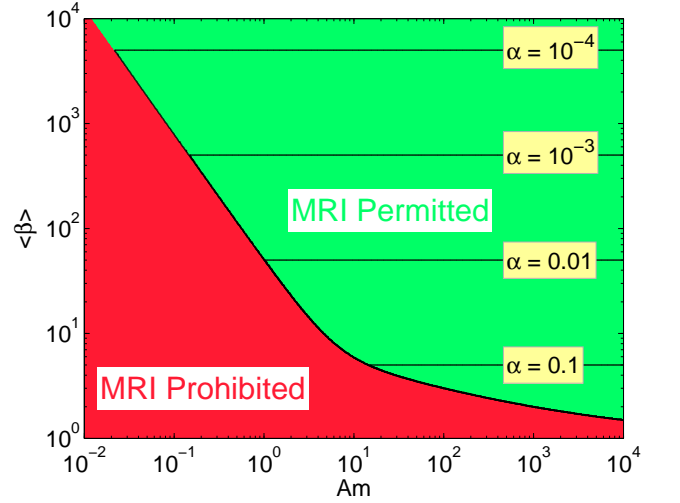}
  \caption{Diagnostics of the MRI in the AD regime. At a given $Am$, MRI is permitted
  when $\langle\beta\rangle\geq\beta_{\rm min}(Am)$ (see equation (\ref{eq:constrain})),
  and in the MRI permitted region, the stress $\alpha$ can be inferred given the field
  strength at the saturated state characterized by $\langle\beta\rangle$.}
  \label{fig:constrain}
\end{figure}

In Figure \ref{fig:alphabeta} we show the scattered plot of $\alpha$ and
$\langle\beta\rangle$ from all our simulations with sustained turbulence with different
field geometries. We see that regardless of the initial field geometry and the value of
$Am$, there is a remarkably tight correlation between the two quantities at the saturated
state of the MRI turbulence. The correlation can be represented by
\begin{equation}
\langle\beta\rangle\approx\frac{1}{2\alpha}\ .\label{eq:alphbeta}
\end{equation}
This result is consistent with findings by \citep{HGB95} in ideal MHD simulations, and
extends it to the non-ideal MHD regime. Analytical study of the saturation of the MRI
with Ohmic resistivity by parasitic modes also predicts similar relations \citep{Pessah10}.
More explicitly, this relation translates to
\begin{equation}
\overline{B^2}\approx4\overline{B_rB_\phi}(1+R)\ ,
\end{equation}
where $R$ is the ratio of Reynolds to Maxwell stress (typically $\approx1/3$ in the ideal
MHD case). This relation implies that the Maxwell stress is approximately a fixed fraction of
the total magnetic energy, a reasonable result if the magnetic field is dominated by turbulent
fluctuations. Only two points appear to deviate from this correlation, which correspond to net
toroidal flux simulations with $Am=3$. We see from Table \ref{tab:turbBy} that in these two
simulations, the magnetic energy is dominated by the background toroidal field (i.e., in the
transition where MRI is marginally sustained), therefore producing smaller
$\langle\beta\rangle$ than predicted.

Next, we consider the relation between $\langle\beta\rangle$ and $Am$.
In Figure \ref{fig:Ambeta} we show scattered plot of $Am$ and $\langle\beta\rangle$. We
see that $\langle\beta\rangle$ does not strongly correlate with $Am$, but also
depends on the field geometry and the initial field strength. However, combining the
simulation from all field geometries allow us to identify the lower bound of
$\langle\beta\rangle$ at a given $Am$, denoted by $\beta_{\rm min}$, below which the
field is too strong for to be destabilized based on our discussions before. For
$Am\leq1$, we have performed simulations with the smallest possible $\beta_0$ such
that the most unstable mode marginally fit into the disk height $H$, and the value of
$\beta_{\rm min}$ identified in this regime is robust. For $Am>1$, the exploration
on $\beta_0$ may not be as complete especially in the simulations with both net vertical
and toroidal fluxes and the actual $\beta_{\rm min}$ may be somewhat smaller than
obtained here. Nevertheless, this regime is closer to ideal MHD and is less concerning.
By combining all the available simulations, we obtain a fitting formula for
$\beta_{\min}$ given by
\begin{equation}
\beta_{\rm min}(Am)=\bigg[\bigg(\frac{50}{Am^{1.2}}\bigg)^2
+\bigg(\frac{8}{Am^{0.3}}+1\bigg)^2\bigg]^{1/2}\ ,\label{eq:constrain}
\end{equation}
and is indicated in Figure \ref{fig:Ambeta}. It asymptotes to $1$ at
$Am\rightarrow\infty$ as one expects, while approaches $50/Am^{1.2}$ for $Am\lesssim1$.

The constraint on $\beta_{\rm min}$ at a given $Am$ allows us to identify the regions
in the $Am$-$\langle\beta\rangle$ plane at which MRI can or can not operate. In
the mean time the correlation between $\alpha$ and $\langle\beta\rangle$ provide the
corresponding stress when MRI is permitted. Combining them together, the main results
from the whole paper are best summarized in Figure \ref{fig:constrain}. MRI permitted regions
are in the upper right with the boundary given by equation (\ref{eq:constrain}). It provides
useful diagnostics on the properties of the MRI in the AD regime in a concise fashion.

First, at a given $Am$, the ultimate strength of the MRI turbulence (e.g., $\alpha$
and $\langle\beta\rangle$) depends on the field geometry (including the net flux), but
there exists a maximum $\alpha$ (or minimum $\langle\beta\rangle$) at the most favorable
field geometry (usually contains both net vertical and toroidal fluxes). One way to think
about it is to starts with a weak regular field as we perform our simulations. As the system
evolves and as the MRI amplifies the field, the corresponding position of the system in the
diagram moves downward and until it stops at some
$\langle\beta\rangle\gtrsim\beta_{\rm min}$.

Second, MRI can be self-sustained for any value of $Am$ even for $Am\ll1$. Although we
have explored the $Am$ parameter down to $Am=0.1$, we believe that it can be
extended to further smaller $Am$ because of the following reasons. Linear analysis
by \citet{KunzBalbus04} and \citet{Desch04} shows in the presence of both vertical and
toroidal field, MRI can grow at appreciable rate (approximately $0.13\Omega^{-1}$ when
$B_\phi/B_z=4$) even in the limit of $Am\rightarrow0^+$ provided that the field is
sufficiently weak. This means that MRI turbulence can always be self-sustained.
Meanwhile, we find that the linear dispersion relation has already approached the small
$Am$ asymptote for $Am\lesssim0.3$. Therefore, we expect the trend in Figure
\ref{fig:Ambeta} on $\beta_{\rm min}$ to hold to further smaller $Am$ values.

Third, the boundary between the MRI permitted and prohibited regions is only suggestive
but it does not necessarily imply sharp transitions. Our simulations are restricted by the
limited box height ($H$) since they are unstratified. In reality, as one increases the field
strength, the transition from sustained MRI turbulence to its suppression involves the effect
of vertical stratification of gas density in the disks, and may be a smooth process. Before
justified by stratified simulations, which is left for our future work, this result should be
taken with some caution. In particular, when vertical stratification is include, linear analysis
by \citet{GammieBalbus94} and \citet{SalmeronWardle05} for ideal and non-ideal MHD
have suggested the existence of global modes in the disk even in low $\beta_0$ and small
Elsasser number. On the other hand, in the case of Ohmic resistivity, the criterion that
Ohmic Elsasser number equals one being the boundary between MRI permitted and
suppressed regions identified in unstratified simulations
\citep{Sano_etal98,Fleming_etal00,SanoStone02b} do agree with results from stratified
simulations \citep{FlemingStone03,Turner_etal07,IlgnerNelson08}.

\section[]{Summary and Discussion}\label{sec:conclusion}

Weakly ionized plasma is subject to a number of non-ideal MHD effects due to
the collisional coupling between the ionized species and the neutrals. Among
them, ambipolar diffusion (AD) results from the relative motion between the ions
and the neutrals. It becomes most important when the gyro-frequencies of both
the electrons and the ions in the magnetic field are larger than their collision
frequencies with the neutrals, so all ionized species are effectively coupled to
the magnetic field. Consequently, AD usually dominates other non-ideal MHD
effects (Ohmic resistivity and Hall effect) in regions with low density
and high magnetic field. When the ion inertia is negligible and when the electron
recombination time is much less than the dynamical time, AD is in the ``strong
coupling" limit, and can be studied by single-fluid models. In this limit, the effect
of AD is fully characterized by the parameter $Am$, the number of
times a neutral molecule/atom collide with the ions in a dynamical time, and is the
equivalence of the Elsasser number for Ohmic resistivity. The effect of AD
becoemes dynamically important when $Am$ approaches order unity.

In weakly ionized disks such as the protoplanetary disks (PPDs), AD dominates
other non-ideal MHD effects in the disk upper layer as well as the outer regions of
the PPDs \citep{Wardle07,Bai11}. The magnetorotational instability (MRI), which
is considered as the major mechanism for providing angular momentum transport via
the MHD turbulence, is strongly affected by AD. In the linear regime, the growth of the
MRI is reduced or
suppressed in the presence of AD, depending on the value of $Am$ and the
magnetic field geometry \citep{BlaesBalbus94,KunzBalbus04,Desch04}. Two-fluid
simulations of the non-linear evolution of the MRI with AD by \citet{HawleyStone98}
showed that significant turbulence and angular momentum transport occurs when
$Am\gtrsim100$. However, they ignored the processes of ionization and
recombination, and the results are not directly applicable to the PPDs, where the
strong coupling limit holds \citep{Bai11}.

We have implemented AD in the strong coupling limit in the Athena MHD code
using a first-order accurate operator-split method. Its performance is verified
by numerical tests on standing C-type shocks and the damping of MHD waves.
We then perform local shearing box numerical simulations to study the effect of
AD on the non-linear evolution of the MRI in the strong coupling limit. Our
simulations are vertically unstratified with vertical box size equals the disk scale
height $H$. The main purpose of this paper is to study how the properties of the
MRI turbulence is affected by AD, especially the condition under which MRI
turbulence can be self-sustained. We perform three groups of simulations with
different magnetic field configurations, and the main results are summarized below.

\begin{enumerate}
\item Net vertical flux simulations. Unstable linear MRI modes always exist for any
value of $Am$, with longer wavelength and smaller growth rate as $Am$ gets smaller.
MRI turbulence can be self-sustained as long as the wavelength of the most unstable
mode $\lambda_m$ is within $H$ (so the vertical field has to be progressively weaker
for $Am\rightarrow0^+$). At fixed $Am$, the total turbulent stress $\alpha$
increases monotonically with net vertical flux, until reaches the maximum when
$\lambda_m\approx H$. The maximum value of $\alpha$ rapidly decreases with
$Am$ when $Am<10$, from about $0.4$ at $Am\rightarrow\infty$ down to about
$0.007$ at $Am=1$. It falls below $10^{-3}$ at $Am\approx0.3$ and is around
of $10^{-4}$ at $Am=0.1$.

\item Net toroidal flux simulations. This field configuration is more stable. At fixed
$Am\gtrsim3$ and net flux, the turbulent stress $\alpha$ from net toroidal flux
simulations is smaller than that from net vertical flux simulations by about an order
of magnitude. We do not find any evidence that MRI turbulence can be self-sustained
at the level of $\alpha\gtrsim10^{-4}$ when $Am\lesssim1$ for any net toroidal flux.

\item Simulations with both net vertical and net toroidal fluxes.
When $Am\gtrsim1$ and fixed net vertical flux, the strength of the MRI turbulence is
similar to the pure net vertical flux case, but slowly increases with the net toroidal flux.
When $Am\lesssim1$, the most unstable mode has non-zero radial wavenumber
comparable to the vertical wavenumber, and the fastest growth rate asymptotes to
some appreciable value even as $Am\rightarrow0^+$. The maximum value of the
turbulent stress $\alpha$ largely exceeds the pure net vertical flux case when
$Am<1$, with $\alpha\approx6\times10^{-4}$ at $Am=0.1$.

\end{enumerate}

In addition, we find that similar to the effect of Ohmic dissipation, the ratio of the
fluctuating part of the magnetic energy density to the kinetic energy density
decreases as AD is stronger. Similarly, the ratio of Maxwell stress to Reynolds
stress also drops at smaller $Am$. The power spectra density of the MRI
turbulence in the AD dominated regime does not show any new features
other than a rescaling from that in the ideal MHD case. We do not find any
evidence that AD leads to the formation of sharp current structures in the MRI
turbulence as proposed by \citet{BrandenburgZweibel94}, but we confirm that AD
tends to reduce the component of current perpendicular to the direction of
magnetic field \citep{Brandenburg_etal95}, although to a lesser extent.

Combining the results from these three groups of simulations, we find a strong
correlation between the turbulent stress $\alpha$ and the gas to magnetic pressure
ratio $\beta$ at the saturated state, given by $\langle\beta\rangle\approx1/2\alpha$.
The sustainability and saturation level of the MRI turbulence depends on the value of
$Am$, the magnetic field geometry, and the magnetic field strength. It is best
summarized in Figure \ref{fig:constrain}. In short, at a given $Am$, there exists a
maximum value of turbulent stress $\alpha$ achievable from the most favorable
geometries (generally with both net vertical and toroidal fluxes). Correspondingly,
at a given $Am$, there exists a maximum field strength above which MRI is suppressed,
and the maximum field strength rapidly decreases with decreasing $Am$. 
For future reference, we quote the turbulent stress $\alpha=7\times10^{-3}$ and
$6\times10^{-4}$ as the maximum value we have found in our simulations at $Am=1$
and $0.1$ respectively.

In principle, in the presence of both net vertical and net toroidal fluxes, MRI turbulence
can be self-sustained at any values of $Am$, provided that the magnetic field is
sufficiently small. Nevertheless, the resulting turbulent stress $\alpha$ would be much
smaller than $6\times10^{-4}$ when $Am<0.1$, and would be inefficient in transporting
angular momentum in most astrophysical disks. Moreover, we recall that AD dominates
other non-ideal MHD effects (Ohmic resistivity and Hall effect) at relatively low density
and relatively large magnetic field. When the magnetic field is too weak as required for
the wavelength of the most unstable MRI mode to be within $H$, AD may no longer be
the dominant non-ideal MHD effect, and our results are not directly applicable. Therefore,
while generalization of our results to $Am<0.1$ are possible, it may not be physically
relevant.

We have shown that the effect of AD on the non-linear evolution of the MRI depends
on the field geometry, which is arbitrarily assigned in our local shearing box
simulations. In reality, the field geometry depends on the global evolution of the
disk. Let us imagine one scenario and apply our local simulation results to a global
picture. If the disk is initially threaded by a very weak vertical field (e.g.,
\citealp{Armitage98}), the initial growth of the MRI may be governed by the Ohmic
resistivity and/or the Hall effect. AD takes over when substantial amplification of the
field occurs and dominate the non-linear evolution of the MRI. The saturation of the MRI
generates strong vertical and azimuthal fields, and redistributes the vertical and
toroidal fluxes across the entire disk. No longer limited by the enforced net fluxes as in
our local simulations, the saturation of the MRI could probably arrange the magnetic
field to achieve the most favorable local field configurations such that the field is
maximumly amplified and $\langle\beta\rangle\approx\beta_{\rm min}$ is achieved.
Therefore, given the value of $Am$ in the disk, and provided that AD dominates other
non-ideal MHD effects, we expect the field strength in the disk to be largely determined
by the value of $\beta_{\rm min}(Am)$, via Equation (\ref{eq:constrain}), from which one
can also obtain a rough estimate of $\alpha\approx1/2\beta_{\rm min}$.

Our results are mostly relevant to the structure and evolution of the PPDs. Details
about the application require considerations of the ionization and recombination
processes in the disks with an appropriate chemistry model, which is beyond the scope
of this paper. In our companion paper \citep{Bai11}, we will take into account all
non-ideal MHD effects together and study their implications in the PPDs.

\acknowledgments

We thank Shane Davis for providing the code for analyzing the power spectrum and Jeremy
Goodman for help discussions. This work is supported by NSF grant AST-0908269. X.-N.B
acknowledges support from NASA Earth and Space Science Fellowship.



\begin{thebibliography}{67}
\expandafter\ifx\csname natexlab\endcsname\relax\def\natexlab#1{#1}\fi

\bibitem[{{Armitage}(1998)}]{Armitage98}
{Armitage}, P.~J. 1998, \apjl, 501, L189

\bibitem[{{Bai}(2011)}]{Bai11}
{Bai}, X. 2011, \apj, in preparation

\bibitem[{{Bai} \& {Goodman}(2009)}]{BaiGoodman09}
{Bai}, X.-N. \& {Goodman}, J. 2009, \apj, 701, 737

\bibitem[{{Balbus}(2003)}]{Balbus03}
{Balbus}, S.~A. 2003, \araa, 41, 555

\bibitem[{{Balbus} \& {Hawley}(1991)}]{BH91}
{Balbus}, S.~A. \& {Hawley}, J.~F. 1991, \apj, 376, 214

\bibitem[{{Balbus} \& {Hawley}(1992)}]{BalbusHawley92}
---. 1992, \apj, 400, 610

\bibitem[{{Balbus} \& {Terquem}(2001)}]{BalbusTerquem01}
{Balbus}, S.~A. \& {Terquem}, C. 2001, \apj, 552, 235

\bibitem[{{Balsara}(1996)}]{Balsara96}
{Balsara}, D.~S. 1996, \apj, 465, 775

\bibitem[{{Blaes} \& {Balbus}(1994)}]{BlaesBalbus94}
{Blaes}, O.~M. \& {Balbus}, S.~A. 1994, \apj, 421, 163

\bibitem[{{Bodo} {et~al.}(2008){Bodo}, {Mignone}, {Cattaneo}, {Rossi}, \&
  {Ferrari}}]{Bodo_etal08}
{Bodo}, G., {Mignone}, A., {Cattaneo}, F., {Rossi}, P., \& {Ferrari}, A. 2008,
  \aap, 487, 1

\bibitem[{{Brandenburg} {et~al.}(1995){Brandenburg}, {Nordlund}, {Stein}, \&
  {Torkelsson}}]{Brandenburg_etal95}
{Brandenburg}, A., {Nordlund}, A., {Stein}, R.~F., \& {Torkelsson}, U. 1995,
  \apj, 446, 741

\bibitem[{{Brandenburg} \& {Zweibel}(1994)}]{BrandenburgZweibel94}
{Brandenburg}, A. \& {Zweibel}, E.~G. 1994, \apjl, 427, L91

\bibitem[{{Chiang}  \& {Murray-Clay}(2007)}]{CMC07} {Chiang}, E., \&
{Murray-Clay}, R.\ 2007, Nature Physics, 3, 604

\bibitem[{{Choi} {et~al.}(2009){Choi}, {Kim}, \& {Wiita}}]{Choi_etal09}
{Choi}, E., {Kim}, J., \& {Wiita}, P.~J. 2009, \apjs, 181, 413

\bibitem[{{Davis} {et~al.}(2010){Davis}, {Stone}, \& {Pessah}}]{Davis_etal10}
{Davis}, S.~W., {Stone}, J.~M., \& {Pessah}, M.~E. 2010, \apj, 713, 52

\bibitem[{{Desch}(2004)}]{Desch04}
{Desch}, S.~J. 2004, \apj, 608, 509

\bibitem[{{Draine}(1980)}]{Draine80}
{Draine}, B.~T. 1980, \apj, 241, 1021

\bibitem[{{Falle}(2003)}]{Falle03}
{Falle}, S.~A.~E.~G. 2003, \mnras, 344, 1210

\bibitem[{{Fleming} \& {Stone}(2003)}]{FlemingStone03}
{Fleming}, T. \& {Stone}, J.~M. 2003, \apj, 585, 908

\bibitem[{{Fleming} {et~al.}(2000){Fleming}, {Stone}, \&
  {Hawley}}]{Fleming_etal00}
{Fleming}, T.~P., {Stone}, J.~M., \& {Hawley}, J.~F. 2000, \apj, 530, 464

\bibitem[{{Fromang} \& {Nelson}(2006)}]{FromangNelson06}
{Fromang}, S. \& {Nelson}, R.~P. 2006, \aap, 457, 343

\bibitem[{{Fromang} \& {Papaloizou}(2007)}]{FromangPap07b}
{Fromang}, S. \& {Papaloizou}, J. 2007, \aap, 476, 1113

\bibitem[{{Gammie}(1996)}]{Gammie96}
{Gammie}, C.~F. 1996, \apj, 457, 355

\bibitem[{{Gammie} \& {Balbus}(1994)}]{GammieBalbus94}
{Gammie}, C.~F. \& {Balbus}, S.~A. 1994, \mnras, 270, 138

\bibitem[{{Gardiner} \& {Stone}(2005)}]{GardinerStone05}
{Gardiner}, T.~A. \& {Stone}, J.~M. 2005, Journal of Computational Physics,
  205, 509

\bibitem[{{Gardiner} \& {Stone}(2008)}]{GardinerStone08}
---. 2008, Journal of Computational Physics, 227, 4123

\bibitem[{{Goldreich} \& {Lynden-Bell}(1965)}]{GoldreichLyndenBell65}
{Goldreich}, P. \& {Lynden-Bell}, D. 1965, \mnras, 130, 125

\bibitem[{{Goodman} \& {Xu}(1994)}]{GoodmanXu94}
{Goodman}, J. \& {Xu}, G. 1994, \apj, 432, 213

\bibitem[{{Hawley}(2000)}]{Hawley00}
{Hawley}, J.~F. 2000, \apj, 528, 462

\bibitem[{{Hawley}(2001)}]{Hawley01}
---. 2001, \apj, 554, 534

\bibitem[{{Hawley} {et~al.}(1995){Hawley}, {Gammie}, \& {Balbus}}]{HGB95}
{Hawley}, J.~F., {Gammie}, C.~F., \& {Balbus}, S.~A. 1995, \apj, 440, 742

\bibitem[{{Hawley} \& {Stone}(1998)}]{HawleyStone98}
{Hawley}, J.~F. \& {Stone}, J.~M. 1998, \apj, 501, 758

\bibitem[{{Ilgner} \& {Nelson}(2008)}]{IlgnerNelson08}
{Ilgner}, M. \& {Nelson}, R.~P. 2008, \aap, 483, 815

\bibitem[{{Jin}(1996)}]{Jin96}
{Jin}, L. 1996, \apj, 457, 798

\bibitem[{{Johnson} {et~al.}(2008){Johnson}, {Guan}, \&
  {Gammie}}]{Johnson_etal08}
{Johnson}, B.~M., {Guan}, X., \& {Gammie}, C.~F. 2008, \apjs, 177, 373

\bibitem[{{Kunz} \& {Balbus}(2004)}]{KunzBalbus04}
{Kunz}, M.~W. \& {Balbus}, S.~A. 2004, \mnras, 348, 355

\bibitem[{{Li} {et~al.}(2006){Li}, {McKee}, \& {Klein}}]{Li_etal06}
{Li}, P.~S., {McKee}, C.~F., \& {Klein}, R.~I. 2006, \apj, 653, 1280

\bibitem[{{Mac Low} {et~al.}(1995){Mac Low}, {Norman}, {Konigl}, \&
  {Wardle}}]{MacLow_etal95}
{Mac Low}, M., {Norman}, M.~L., {Konigl}, A., \& {Wardle}, M. 1995, \apj, 442,
  726

\bibitem[{{Masset}(2000)}]{FARGO}
{Masset}, F. 2000, \aaps, 141, 165

\bibitem[{{Miller} \& {Stone}(2000)}]{MillerStone00}
{Miller}, K.~A. \& {Stone}, J.~M. 2000, \apj, 534, 398

\bibitem[{{Oishi} \& {Mac Low}(2009)}]{OishiMacLow09}
{Oishi}, J.~S. \& {Mac Low}, M. 2009, \apj, 704, 1239

\bibitem[{{O'Sullivan} \& {Downes}(2006)}]{OSullivanDownes06}
{O'Sullivan}, S. \& {Downes}, T.~P. 2006, \mnras, 366, 1329

\bibitem[{{O'Sullivan} \& {Downes}(2007)}]{OSullivanDownes07}
---. 2007, \mnras, 376, 1648

\bibitem[{{Papaloizou} \& {Terquem}(1997)}]{PapTerquem97}
{Papaloizou}, J.~C.~B. \& {Terquem}, C. 1997, \mnras, 287, 771

\bibitem[{{Pessah}(2010)}]{Pessah10}
{Pessah}, M.~E. 2010, \apj, 716, 1012

\bibitem[{{Pessah} \& {Chan}(2008)}]{PessahChan08}
{Pessah}, M.~E. \& {Chan}, C. 2008, \apj, 684, 498

\bibitem[{{Pessah} \& {Goodman}(2009)}]{PessahGoodman09}
{Pessah}, M.~E. \& {Goodman}, J. 2009, \apjl, 698, L72

\bibitem[{{Salmeron} \& {Wardle}(2005)}]{SalmeronWardle05}
{Salmeron}, R. \& {Wardle}, M. 2005, \mnras, 361, 45

\bibitem[{{Sano} {et~al.}(1998){Sano}, {Inutsuka}, \& {Miyama}}]{Sano_etal98}
{Sano}, T., {Inutsuka}, S., \& {Miyama}, S.~M. 1998, \apjl, 506, L57

\bibitem[{{Sano} \& {Stone}(2002{\natexlab{a}})}]{SanoStone02a}
{Sano}, T. \& {Stone}, J.~M. 2002{\natexlab{a}}, \apj, 570, 314

\bibitem[{{Sano} \& {Stone}(2002{\natexlab{b}})}]{SanoStone02b}
---. 2002{\natexlab{b}}, \apj, 577, 534

\bibitem[{{Shakura} \& {Sunyaev}(1973)}]{ShakuraSunyaev73}
{Shakura}, N.~I. \& {Sunyaev}, R.~A. 1973, \aap, 24, 337

\bibitem[{{Shen} {et~al.}(2006){Shen}, {Stone}, \& {Gardiner}}]{ShenStone06}
{Shen}, Y., {Stone}, J.~M., \& {Gardiner}, T.~A. 2006, \apj, 653, 513

\bibitem[{{Shu}(1991)}]{Shu91}
{Shu}, F. 1991, {Physics of Astrophysics, Vol. II: Gas Dynamics}, ed. {Shu, F.}
  (University Science Books)

\bibitem[{{Simon} \& {Hawley}(2009)}]{SimonHawley09}
{Simon}, J.~B. \& {Hawley}, J.~F. 2009, \apj, 707, 833

\bibitem[{{Smith} \& {Mac Low}(1997)}]{SmithMacLow97}
{Smith}, M.~D. \& {Mac Low}, M. 1997, \aap, 326, 801

\bibitem[{{Stone}(1997)}]{Stone97}
{Stone}, J.~M. 1997, \apj, 487, 271

\bibitem[{{Stone} {et~al.}(2000){Stone}, {Gammie}, {Balbus}, \&
  {Hawley}}]{Stone_etal00}
{Stone}, J.~M., {Gammie}, C.~F., {Balbus}, S.~A., \& {Hawley}, J.~F. 2000,
  Protostars and Planets IV, 589

\bibitem[{{Stone} \& {Gardiner}(2010)}]{StoneGardiner10}
{Stone}, J.~M. \& {Gardiner}, T.~A. 2010, \apjs, 189, 142

\bibitem[{{Stone} {et~al.}(2008){Stone}, {Gardiner}, {Teuben}, {Hawley}, \&
  {Simon}}]{Stone_etal08}
{Stone}, J.~M., {Gardiner}, T.~A., {Teuben}, P., {Hawley}, J.~F., \& {Simon},
  J.~B. 2008, \apjs, 178, 137

\bibitem[{{Stone} {et~al.}(1996){Stone}, {Hawley}, {Gammie}, \&
  {Balbus}}]{SHGB96}
{Stone}, J.~M., {Hawley}, J.~F., {Gammie}, C.~F., \& {Balbus}, S.~A. 1996,
  \apj, 463, 656

\bibitem[{{Tilley} \& {Balsara}(2008)}]{TilleyBalsara08}
{Tilley}, D.~A. \& {Balsara}, D.~S. 2008, \mnras, 389, 1058

\bibitem[{{Toth}(1994)}]{Toth94}
{Toth}, G. 1994, \apj, 425, 171

\bibitem[{{Turner} \& {Sano}(2008)}]{TurnerSano08}
{Turner}, N.~J. \& {Sano}, T. 2008, \apjl, 679, L131

\bibitem[{{Turner} {et~al.}(2007){Turner}, {Sano}, \&
  {Dziourkevitch}}]{Turner_etal07}
{Turner}, N.~J., {Sano}, T., \& {Dziourkevitch}, N. 2007, \apj, 659, 729

\bibitem[{{Wardle}(1999)}]{Wardle99}
{Wardle}, M. 1999, \mnras, 307, 849

\bibitem[{{Wardle}(2007)}]{Wardle07}
---. 2007, \apss, 311, 35

\bibitem[{{Winters} {et~al.}(2003){Winters}, {Balbus}, \&
  {Hawley}}]{Winters_etal03}
{Winters}, W.~F., {Balbus}, S.~A., \& {Hawley}, J.~F. 2003, \mnras, 340, 519

\end{thebibliography}

\label{lastpage}
\end{document}